\documentclass[pra,letterpaper,eqsecnum,preprint]{revtex4}%
\usepackage{amsfonts}
\usepackage{amsmath}
\usepackage{amssymb}
\usepackage{graphicx}%
\newtheorem{theorem}{Theorem}

\begin{document}
\title{Hyperspherical Description of the Degenerate Fermi Gas: S-wave Interactions.}
\author{Seth T. Rittenhouse$^{1}$}
\author{M. J. Cavagnero$^{2}$}
\author{Javier von Stecher$^{1}$}
\author{Chris H. Greene$^{1}$}
\affiliation{$^{1}$Department of Physics and JILA, University of Colorado, Boulder,
Colorado 80309-0440\\
$^{2}$Department of Physics and Astronomy, University of Kentucky, Lexington,
Kentucky 40506-0055}
\date{\today}
\begin{abstract}
We present a unique theoretical description of the physics of the spherically
trapped $N$-atom degenerate Fermi gas (DFG) at zero temperature based on an
ordinary Schr\"{o}dinger equation with a microscopic, two body interaction
potential. With a careful choice of coordinates and a variational
wavefunction, the many body Schr\"{o}dinger equation can be accurately
described by a \emph{linear}, one dimensional effective Schr\"{o}dinger
equation in a single collective coordinate, the rms radius of the gas.
Comparisons of the energy, rms radius and peak density of ground state energy
are made to those predicted by Hartree-Fock (HF). Also the lowest radial
excitation frequency (the breathing mode frequency) agrees with a sum rule
calculation, but deviates from a HF prediction.

\end{abstract}
\maketitle

\section{Introduction}

The realization of the degenerate fermi gas (DFG) in a dilute gas of fermionic
atoms has triggered widespread interest in the nature of these systems. This
achievement combined with the use of a Feshbach resonance allows for a quantum
laboratory in which many quantum phenomena can be explored over a wide range
of interaction strengths. This leads to a large array of complex behaviors
including the discovery of highly correlated BCS-like pairing for effectively
attractive interactions
\cite{regal2004orc,zwierlein2004cpf,bartenstein2004cmb,bourdel2004esb,kinast2004esr}%
. While these pairing phenomena are extremely interesting, we believe that the
physics of a pure degenerate Fermi gas, in which the majority of atoms
successively fill single particle states, is worthy of further study even in
the absence of pairing. This topic is addressed in the present paper, which
follows a less complete archived study of the topic.\cite{condmat}

The starting point for this study is that of a hyperspherical treatment of the
problem in which the gas is described by a set of $3N-1$ angular coordinates
on the surface of a $3N$ dimensional hypersphere of radius $R$. Here $N$ is
the number of atoms in the system. This formulation is inspired by a similar
study of the Bose-Einstein
condensate.\cite{bohn_esry_greene_hsbec,kim2000elt,kushibe2004aha} Furthermore
these same coordinates have been applied to finite
nuclei.\cite{SmirnovShitikova} The formulation is a rigorous variational
treatment of a many-body Hamiltonian, aside from the limitations of the
assumption of pairwise, zero-range interactions. In this paper, we consider
only filled energy shells of atoms. This is done for analytic and
calculational simplicity, but the treatment should apply to any number of
atoms in open shells with modest extensions that are discussed briefly.

The main goal of this study is to describe the motion of the gas in a
\emph{single }collective coordinate $R$, which describes the overall extent of
the gas. The benefit of this strategy is that the behavior of the gas is
reduced to a single one-dimensional \emph{linear} Schr\"{o}dinger equation
with an effective hyperradial potential. The use of a real potential then
lends itself to the intuitive understanding of normal Schr\"{o}dinger quantum
mechanics. This method also allows for the calculation of physical quantities
such as the energy and rms radius of the ground state; these observables agree
quantitatively with those computed using Hartree-Fock methods. The method also
yields a visceral understanding of a low energy collective oscillation of the
gas, i.e. a breathing mode.

Beyond the intuitive benefits of reducing the problem to an effective
one-dimensional Schr\"{o}dinger equation, another motivation for developing
this hyperspherical viewpoint is that it has proven effective in other
contexts for describing processes involving fragmentation or collisions in
few-body and many-body systems.\cite{esry1999rta,blume00} Such processes would
be challenging to formulate using field theory or RPA or configuration
interaction viewpoints, but they emerge naturally and intuitively, once the
techniques for computing the hyperspherical potential curves for such systems
are adequately developed. To this end we view it as a first essential step, in
the development of a more comprehensive theory, to calculate the ground-state
and low-lying excited state properties within this framework. Then we can
ascertain whether the hyperspherical formulation is capable of reproducing the
key results of other, more conventional descriptions, which start instead from
a mean-field theory perspective.

The paper is organized as follows: Section II develops the formulation that
yields a hyperradial 1D effective Hamiltonian; in Section III we apply this
formalism to a zero-range s-wave interaction and find the effective
Hamiltonian in both the finite $N$ and large $N$ limits. In Sections IIIa and
IIIb we examine the nature of the resulting effective potential for a wide
range of interaction strengths and give comparisons with other known methods,
mainly Hartree-Fock (HF); Section IIIc is a brief discussion of the
simplifications that can be made in the limit where $N\rightarrow\infty$;
finally in Section IV we summarize the results and discuss future avenues of study.

\section{Formulation}

The formalism is similar to that of reference \cite{bohn_esry_greene_hsbec},
but we will reiterate it for clarity and to make this article self-contained.
Consider a collection of $N$ identical fermionic atoms of mass $m$ in a
spherically symmetric trap with oscillator frequency $\omega$, distributed
equally between two internal spin substates. The governing Hamiltonian is%
\begin{equation}
H=\dfrac{-\hbar^{2}}{2m}\sum_{i=1}^{N}\nabla_{i}^{2}+\dfrac{1}{2}m\omega
^{2}\sum_{i=1}^{N}r_{i}^{2}+\sum_{i>j}U_{int}\left(  \vec{r}_{ij}\right)
\end{equation}
where $U_{int}\left(  \vec{r}\right)  $ is an arbitrary two-body interaction
potential and $\vec{r}_{ij}=\vec{r}_{i}-\vec{r}_{j}$. We ignore interaction
terms involving three or more bodies. In general, the Schr\"{o}dinger equation
that comes from this Hamiltonian is very difficult to solve. Our goal is to
simplify the system by describing its behavior in terms of a \emph{single
collective coordinate}. To achieve this aim we transform this Hamiltonian into
a set of collective hyperspherical coordinates; the hyperradius, $R$, of this
set is given by the root mean square distance of the atoms from the center of
the trap.%
\begin{equation}
R\equiv\left(  \dfrac{1}{N}\sum_{i=1}^{N}r_{i}^{2}\right)  ^{1/2}.
\label{hyperr}%
\end{equation}
So far we only have one coordinate for the system, but we need to account for
all $3N$ spatial coordinates and also the spin degrees of freedom. This leaves
$3N-1$ angular coordinates left to define. We have $2N$ of the angles as the
independent particle spherical polar coordinate angles $\left(  \theta
_{1},\phi_{1},\theta_{2},\phi_{2},...,\theta_{N},\phi_{N}\right)  $. The
remaining $N-1$ hyperangles are chosen by the convention used in
\cite{SmirnovShitikova}, which describes correlated motions in the radial
distances of the atoms from the trap center,%
\begin{align}
\tan\alpha_{i}  &  =\dfrac{\sqrt{\sum_{j=1}^{i}r_{j}^{2}}}{r_{i+1}%
},\label{hypera}\\
i  &  =1,2,3,...,N-1.\nonumber
\end{align}
Alternatively we may write this as%
\begin{align}
r_{n}  &  =\sqrt{N}R\cos\alpha_{n-1}\prod\limits_{j=n}^{N-1}\sin\alpha
_{j}\label{hypera2}\\
0  &  \leq\alpha_{j}\leq\dfrac{\pi}{2},\text{ }j=1,2,...,N-1\nonumber
\end{align}
where we define $\cos\alpha_{0}\equiv1$ and $\prod\limits_{j=N}^{N-1}%
\sin\alpha_{j}\equiv1$. For the purposes of this study, the set of $3N-1$
hyperangles will be referred to collectively as $\Omega$. The particular
definition of the hyperangles will not play a significant role in the actual
formalism, but we give the definitions here for completeness.

After carrying out this coordinate transformation on the sum of Laplacians,
the kinetic energy becomes \cite{Avery}%
\begin{equation}
\dfrac{-\hbar^{2}}{2M}\left[  \dfrac{1}{R^{3N-1}}\dfrac{\partial}{\partial
R}\left(  R^{3N-1}\dfrac{\partial}{\partial R}\right)  -\dfrac{\mathbf{\Lambda
}^{2}}{R^{2}}\right]  . \label{kineticE}%
\end{equation}
\qquad Here $M=Nm$ and $\hbar\mathbf{\Lambda}$ is the\ grand angular momentum
operator which is similar to the conventional angular momentum operator and is
defined by%
\begin{equation}
\mathbf{\Lambda}^{2}=-\sum_{i>j}\Lambda_{ij}^{2},\Lambda_{ij}=x_{i}%
\dfrac{\partial}{\partial x_{j}}-x_{j}\dfrac{\partial}{\partial x_{i}}
\label{hypermom}%
\end{equation}
for all cartesian components, $x_{i}$, of the $3N$ dimensional space. The
isotropic spherical oscillator potential becomes simply%
\begin{equation}
\dfrac{1}{2}m\omega^{2}\sum_{i=1}^{N}r_{i}^{2}=\dfrac{1}{2}M\omega^{2}R^{2}.
\label{oscillatorpot}%
\end{equation}
The sum of Eqs. \ref{kineticE} and \ref{oscillatorpot} gives a
time-independent Schr\"{o}dinger equation, $H\Psi\left(  R,\Omega\right)
=E\Psi\left(  R,\Omega\right)  $, of the form
\begin{align}
0=  &  \left[  \dfrac{-\hbar^{2}}{2M}\left(  \dfrac{\partial^{2}}{\partial
R^{2}}+\dfrac{\left(  3N-1\right)  \left(  3N-3\right)  }{2R^{2}}%
-\dfrac{\mathbf{\Lambda}^{2}}{R^{2}}\right)  +\dfrac{1}{2}M\omega^{2}%
R^{2}\right. \\
&  +\left.  \sum_{i>j}U_{int}\left(  \vec{r}_{i}-\vec{r}_{j}\right)
-E\right]  R^{\left(  3N-1\right)  /2}\Psi\left(  R,\Omega\right) \nonumber
\end{align}
where $\Psi\left(  R,\Omega\right)  $ has been multiplied by $R^{\left(
3N-1\right)  /2}$ to remove first derivative terms in the hyperradius. These
mathematical transformations have not yet accomplished much. We started with a
$3N$ dimensional Schr\"{o}dinger equation to solve, and we still have a $3N$
dimensional Schr\"{o}dinger equation. To simplify, we assume that $\Psi$ is
approximately separable into an eigenfunction of the operator $\mathbf{\Lambda
}^{2}$, a \textquotedblleft hyperspherical harmonic", multiplied by an unknown
hyperradial function. Hyperspherical harmonics (HHs, see \cite{Avery} for more
details) are generally expressed as products of Jacobi polynomials for any
number of dimensions. Their eigenvalue equation is
\begin{equation}
\mathbf{\Lambda}^{2}\Phi_{\lambda\nu}\left(  \Omega\right)  =\lambda\left(
+\lambda+3N-1\right)  \Phi_{\lambda\nu}\left(  \Omega\right)  \label{HHeval}%
\end{equation}
Where $\lambda=0,1,2,...$ and $\nu$ (omitted below for brevity) stands for the
$3N-2$ other quantum numbers that are needed to distinguish between the
(usually quite large) degeneracies for a given $\lambda$. The separability
ansatz implies that%
\begin{equation}
\Psi\left(  R,\Omega\right)  =F\left(  R\right)  \Phi_{\lambda}\left(
\Omega\right)  .
\end{equation}
Here we assume that $\Phi_{\lambda}\left(  \Omega\right)  $ is the HH that
corresponds to the lowest value of the hyperangular momentum that is allowed
for the given symmetry of the problem, i.e. that is antisymmetric with respect
to interchange of indistinguishable fermions. This choice of trial
wavefunction indicates that we expect the overall energetics of the gas to be
described by its size; as such, fixing the hyperangular behavior is equivalent
with fixing the configuration of the atoms in the gas. In nuclear physics this
is known as the K harmonic method. Alternatively, this may be viewed as
choosing a trial wavefunction whose hyperradial behavior will be variationally
optimized, but whose hyperangular behavior is that of a non-interacting trap
dominated gas of fermions.

To utilize this as a trial wavefunction, we evaluate the expectation value
$\left\langle \Phi_{\lambda}\left\vert H\right\vert \Phi_{\lambda
}\right\rangle $, where the integration is taken over all hyperangles at a
fixed hyperradius. This approach gives a new effective linear 1D
Schr\"{o}dinger equation $H_{eff}R^{\left(  3N-1\right)  /2}F\left(  R\right)
=ER^{\left(  3N-1\right)  /2}F\left(  R\right)  $ in terms of an effective
Hamiltonian $H_{eff}$ given by
\begin{equation}
\dfrac{-\hbar^{2}}{2M}\left(  \dfrac{d^{2}}{dR^{2}}-\dfrac{K\left(
K+1\right)  }{R^{2}}\right)  +\dfrac{1}{2}M\omega^{2}R^{2}+\sum_{i>j}%
\left\langle \Phi_{\lambda}\left\vert U_{int}\left(  \vec{r}_{ij}\right)
\right\vert \Phi_{\lambda}\right\rangle .
\end{equation}
Here $K=\lambda+3\left(  N-1\right)  /2$.

We now must force our trial wavefunction to obey the antisymmetry condition of
fermionic atoms. To antisymmetrize the total wave function $F\left(  R\right)
\Phi_{\lambda}\left(  \Omega\right)  $ note that Eq. \ref{hyperr} indicates
that $R$ is completely symmetric under all particle coordinate exchange, thus
the antisymmetrization of the wavefunction must only affect $\Phi_{\lambda
}\left(  \Omega\right)  $. Finding a completely antisymmetric $\Phi_{\lambda
}\left(  \Omega\right)  $ for any given $\lambda$ is generally quite difficult
and is often done using recursive techniques like coefficient of fractional
parentage expansions (see \cite{SmirnovShitikova, Cavagnero86,barnea1999rmc}
for more details) or using a basis of Slater determinants of independent
particle wave functions.\cite{Timofeyuk02,Timofeyuk04,Fabre78,Fabre05} We use
a simplified version of the second method combined with the following theorem,
proved in reference \cite{Cavagnero86} and developed in Appendix B.

\begin{theorem}
The ground state of any non-interacting set of $N$ particles in an isotropic
oscillator is an eigenfunction of $\mathbf{\Lambda}^{2}$ with minimal
eigenvalue $\lambda\left(  \lambda+3N-2\right)  $ where $\lambda$ is given by
the total number of oscillator quanta in the non-interacting system.%
\begin{equation}
\lambda=\dfrac{E_{NI}}{\hbar\omega}-\dfrac{3N}{2} \label{lambda}%
\end{equation}
where $E_{NI}$ is the total ground state energy of the non-interacting
$N$-body system. \label{theorem1}
\end{theorem}

With Theorem 1 in hand we may find $\Phi_{\lambda}\left(  \Omega\right)  $ in
terms of the independent particle coordinates. The non-interacting ground
state is given by a Slater determinant of single particle solutions.%
\begin{equation}
\Psi_{NI}\left(  \vec{r}_{1},\vec{r}_{2},...,\vec{r}_{N},\sigma_{1},\sigma
_{2},...,\sigma_{N}\right)  =\sum\limits_{P}\left(  -1\right)  ^{p}%
P\prod\limits_{i=1}^{N}R_{n_{i}\ell_{i}}\left(  r_{i}\right)  y_{\ell_{i}%
m_{i}}\left(  \omega_{i}\right)  \left\vert m_{s_{i}}\right\rangle
.\label{Slater1}%
\end{equation}
Here $\sigma_{i}$ is the spin coordinate for the $i$th atom, $R_{n_{i}\ell
_{i}}\left(  r_{i}\right)  $ is the radial solution to the independent
particle harmonic oscillator for the $i$th particle\textbf{ }given by
$rR_{n\ell}\left(  r\right)  =N_{n\ell}\exp\left(  -r^{2}/2l^{2}\right)
\left(  r/l\right)  ^{\ell+1}L_{n}^{\ell+1/2}\left[  \left(  r/l\right)
^{2}\right]  $ where $L_{n}^{\alpha}\left(  r\right)  $ is an associated
Laguerre polynomial with $l=\sqrt{\hbar/m\omega}$. $y_{\ell_{i}m_{i}}\left(
\omega\right)  $ is an ordinary 3D spherical harmonic with $\omega_{i}$ as the
spatial solid angle for the $i$th particle, $\left\vert m_{s_{i}}\right\rangle
$ is a spin ket that will allow for two spin species of atoms, $\left\vert
\uparrow\right\rangle $ and $\left\vert \downarrow\right\rangle $. The sum in
Eq. \ref{Slater1} runs over all possible permutations $P$ of the $N$ spatial
and spin coordinates in the product wavefunction. We now apply Theorem 1 which
directly leads to a $\Psi_{NI}$ that is separable into a hyperangular piece
and a hyperradial piece%
\begin{equation}
\Psi_{NI}\left(  \vec{r}_{1},\vec{r}_{2},...,\vec{r}_{N},\sigma_{1},\sigma
_{2},...\sigma_{N}\right)  =G\left(  R\right)  \Phi_{\lambda}\left(
\Omega,\sigma_{1},\sigma_{2},...\sigma_{N}\right)  .\label{TotalNIwf}%
\end{equation}
Here $G\left(  R\right)  $ is a the nodeless hyperradial wave function
describing the ground state of $N$ non-interacting atoms in an isotropic
oscillator trap. A derivation of $G\left(  R\right)  $ is given in Appendix A:%
\begin{equation}
R^{\left(  3N-1\right)  /2}G\left(  R\right)  =A_{\lambda}\exp\left(
-R^{2}/2\mathcal{L}^{2}\right)  \left(  \dfrac{R}{\mathcal{L}}\right)
^{\lambda+3N/2-1/2},\label{NIhyperrad}%
\end{equation}
here $A_{\lambda}$ is a normalization constant and $\mathcal{L}=\sqrt
{\hbar/M\omega}=l/\sqrt{N}.$ Combining Eq. \ref{TotalNIwf} with Eq.
\ref{NIhyperrad} now gives us $\Phi_{\lambda}\left(  \Omega\right)  $%
\begin{equation}
\Phi_{\lambda}\left(  \Omega,\sigma_{1},\sigma_{2},...\sigma_{N}\right)
=\dfrac{\sum\limits_{P}\left(  -1\right)  ^{p}P\prod\limits_{i=1}^{N}%
R_{n_{i}\ell_{i}}\left(  r_{i}\right)  y_{\ell_{i}m_{i}}\left(  \omega
_{i}\right)  \left\vert m_{s_{i}}\right\rangle }{G\left(  R\right)
}\label{phi}%
\end{equation}
where we must make the variable substitutions in the numerator using Eqs.
\ref{hyperr} and \ref{hypera}. In the following, for brevity, the spin
coordinates $\left(  \sigma_{1},...,\sigma_{N}\right)  $ will be suppressed,
i.e. $\Phi_{\lambda}\left(  \Omega\right)  =\Phi_{\lambda}\left(
\Omega,\sigma_{1},\sigma_{2},...\sigma_{N}\right)  $. Interestingly, this must
be a function only of the hyperangles and thus all of the hyperradial
dependence must cancel out in the right hand side of \ref{phi}. We are now
ready to start calculating the interaction matrix element $\left\langle
\Phi_{\lambda}\left\vert U_{int}\right\vert \Phi_{\lambda}\right\rangle $.

\section{Zero-range s-wave interaction.}

Here we specify $U_{int}\left(  \vec{r}\right)  $ as a zero-range two body
interaction with an interaction strength given by a constant parameter $g$.
\begin{equation}
U_{int}\left(  \vec{r}\right)  =g\delta^{3}\left(  \vec{r}\right)
\label{Swaveint}%
\end{equation}
For a small two body, s-wave scattering length $a$ we know that $g=\dfrac
{4\pi\hbar^{2}a}{m}$.\cite{fermi1934} For stronger interactions,
i.e. $\left\vert k_{f}a\right\vert >1$, this approximation no longer holds and
$g$ must be renormalized.

Now we must calculate the effective interaction matrix element given by%
\begin{equation}
U_{eff}\left(  R\right)  =g\sum_{i>j}\left\langle \Phi_{\lambda}\left\vert
\delta^{3}\left(  \vec{r}_{ij}\right)  \right\vert \Phi_{\lambda}\right\rangle
.\label{Uefftot}%
\end{equation}

\begin{figure}[h]
\begin{center}
\includegraphics[width=3in]{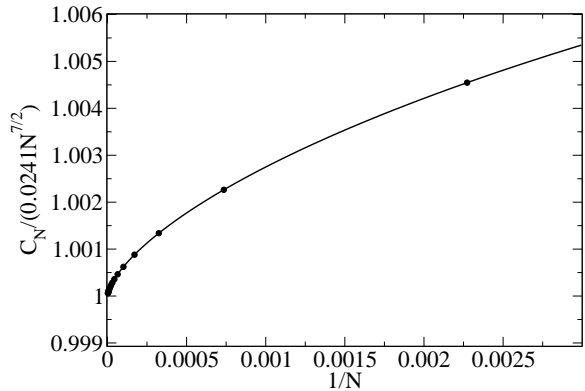}
\end{center}
\caption{Values of $C_{N}$ divided by the large $N$ limit, $C_{N}%
\rightarrow0.02408N^{7/2}$, versus $1/N$ are shown. The circles are the
calculated value while the curve is the fit stated in Eq. \ref{CNlargeN}}%
\label{CNfig}%
\end{figure}

Degenerate ground states cause some complications for this formulation which
we avoid by restricting ourselves to the non-degenerate ground states that
correspond to filled energy shells of the oscillator, i.e. \textquotedblleft
magic numbers" of particles. With moderate extensions the degeneracies can be
taken into account by creating an interaction matrix, but the magic number
restriction should still give a good description of the general behavior of
systems with large numbers of atoms. The total number of atoms and the
hyperangular momentum quantum number $\lambda$ are most conveniently expressed
in terms of the number $n$ of single particle orbital energies filled:%
\begin{subequations}
\begin{align}
N  &  =\dfrac{n\left(  n+1\right)  \left(  n+2\right)  }{3}\label{numpart}\\
\lambda &  =\dfrac{\left(  n-1\right)  n\left(  n+1\right)  \left(
n+2\right)  }{4}\label{Lambda}\\
k_{f}  &  =\sqrt{\dfrac{2m\omega}{\hbar}\left(  n+\dfrac{1}{2}\right)  },
\end{align}
where $k_{f}$ is the peak non-interacting Fermi wave number. In the limit
where $N\gg1$, we write $\lambda$ and $k_{f}$ in terms of the total number of
particles $N,$%
\end{subequations}
\begin{subequations}
\begin{align}
\lambda &  \rightarrow\dfrac{\left(  3N\right)  ^{4/3}}{4}\label{lambdavsN}\\
k_{f}  &  \rightarrow\sqrt{\dfrac{2m\omega}{\hbar}\left(  3N\right)  ^{1/3}}
\label{kfvsN}%
\end{align}
Next we combine \ref{lambdavsN} with $\Phi_{\lambda}\left(  \Omega\right)  $
from \ref{phi} and calculate the interaction matrix element.

Since $\Phi_{\lambda}\left(  \Omega\right)  $ is antisymmetric under particle
exchange we may do a coordinate transposition in the sum $\vec{r}%
_{i}\rightarrow\vec{r}_{2}$ and $\vec{r}_{j}\rightarrow\vec{r}_{1}$. Each
transposition pulls out a negative sign from $\Phi_{\lambda}$ and we are left
with%
\end{subequations}
\begin{align*}
U_{eff}\left(  R\right)   &  =g\sum_{i>j}\left\langle \Phi_{\lambda}\left\vert
\delta^{3}\left(  \vec{r}_{21}\right)  \right\vert \Phi_{\lambda}\right\rangle
\\
&  =g\dfrac{N\left(  N-1\right)  }{2}\left\langle \Phi_{\lambda}\left\vert
\delta^{3}\left(  \vec{r}_{21}\right)  \right\vert \Phi_{\lambda}\right\rangle
.
\end{align*}
Appendix C details the calculation of the matrix element $\left\langle
\Phi_{\lambda}\left\vert \delta^{3}\left(  \vec{r}_{21}\right)  \right\vert
\Phi_{\lambda}\right\rangle $. The result is%
\begin{equation}
U_{eff}\left(  R\right)  =g\dfrac{C_{N}}{N^{3/2}R^{3}}\label{Ueff}%
\end{equation}
where $C_{N}$ is a constant that is dependent only on the number of atoms in
the system. While $C_{N}$ has some complex behavior for smaller numbers of
particles, we have seen that for $N\gtrsim100$, $C_{N}$ quickly converges to%
\begin{align}
C_{N} &  \rightarrow\sqrt{\dfrac{2}{3}}\dfrac{32N^{7/2}}{35\pi^{3}}\left(
1+\dfrac{0.049}{N^{2/3}}-\dfrac{0.277}{N^{4/3}}+...\right)  \label{CNlargeN}\\
&  =0.02408N^{7/2}\left(  1+\dfrac{0.049}{N^{2/3}}-\dfrac{0.277}{N^{4/3}%
}+...\right)  \nonumber
\end{align}
where the higher order terms in $1/N$ were found by fitting a curve to
numerically calculated data. Values of $C_{N}/N^{7/2}$ are shown in Table
\ref{Table1} for several filled shells, Fig. \ref{CNfig} shows both the calculated values of $C_{N}$ and the values from
the fit in Eq. \ref{CNlargeN} versus $1/N.$ In both the table and the plot we
can clearly see the convergence to the large $N$ value of $C_{N}%
\approx0.02408N^{7/2}$.\begin{table}[hptbh]%
\begin{tabular}
[c]{|c|c|c|c|}\hline
$n$ & $N$ & $\lambda$ & $C_{N}/N^{7/2}$\\\hline
$1$ & $2$ & $0$ & $\dfrac{1}{8\pi^{2}}\approx0.0127$\\
$2$ & $8$ & $6$ & $0.0637$\\
$3$ & $20$ & $30$ & $0.0251$\\
$4$ & $40$ & $90$ & $0.0244$\\
$5$ & $70$ & $210$ & $0.02435$\\
$\ldots$ & $\ldots$ & $\ldots$ & $\ldots$\\
$15$ & $1360$ & $14280$ & $0.0241$\\
$30$ & $9920$ & $215760$ & $0.02409$\\
$100$ & $343400$ & $25497450$ & $0.02408$\\\hline
\end{tabular}
\caption{$N$, $\lambda$ and $C_{N}/N^{7/2}$ for the several filled shells. We
can see that $C_{N}/N^{7/2}$ quickly converges to the Thomas-Fermi limiting
value of $32\sqrt{2/3}/35\pi^{3}\approx0.02408$ to several digits.}%
\label{Table1}%
\end{table}

We now may write the effective one dimensional, hyperradial Schr\"{o}dinger
equation, $H_{eff}R^{\left(  3N-1\right)  /2}F\left(  R\right)  =ER^{\left(
3N-1\right)  /2}F\left(  R\right)  $ where $H_{eff}$ is given by%
\begin{equation}
H_{eff}=-\dfrac{\hbar^{2}}{2M}\dfrac{d^{2}}{dR^{2}}+V_{eff}\left(  R\right)
\label{HypersRH}%
\end{equation}
where $V_{eff}\left(  R\right)  $ is an effective hyperradial potential given
by%
\begin{equation}
V_{eff}\left(  R\right)  =\dfrac{\hbar^{2}}{2M}\dfrac{K\left(  K+1\right)
}{R^{2}}+\dfrac{1}{2}M\omega^{2}R^{2}+g\dfrac{C_{N}}{N^{3/2}R^{3}%
}\label{swavepot}%
\end{equation}
We reinforce the idea that this effective Hamiltonian describes the fully
correlated motion of all of the atoms in the trap albeit within the
aforementioned approximations. As expected, for $N=1$, with $C_{1}=0$, Eq.
\ref{swavepot} reduces to a single particle in a trap with $K$ as the angular
momentum quantum number $\ell$. Note that the form of $V_{eff}$ is very
similar to the effective potential found for bosons by the authors of
\cite{bohn_esry_greene_hsbec}. What may be surprising is the extra term of
$3\left(  N-1\right)  /2$ contained in $K$. The kinetic energy term in
$V_{eff}$ is controlled by the hyperangular momentum, which in turn reflects
the total nodal structure of the $N$-fermion wavefunction. This added piece of
hyperangular momentum summarizes the energy cost of confining $N$ fermions in
the trap. This repulsive barrier stabilizes the gas against collapse for
attractive interactions, i.e. $g<0$.

A final transformation simplifies the radial Schr\"{o}dinger equation, namely
setting $E=E_{NI}E^{\prime}$ and $R=\sqrt{\left\langle R^{2}\right\rangle
_{NI}}R^{\prime}$ where
\begin{subequations}
\begin{align}
E_{NI}  &  =\hbar\omega\left(  \lambda+\dfrac{3N}{2}\right)  ,
\label{NIenergy}\\
\left\langle R^{2}\right\rangle _{NI}  &  =l^{2}\left(  \dfrac{\lambda}%
{N}+\dfrac{3}{2}\right)  \label{NIR}%
\end{align}
are the non-interacting expectation values of the energy and squared
hyperradius in the ground state. In the following any hyperradius with a
prime, $R^{\prime}$ denotes the hyperradius in units of $\sqrt{\left\langle
R^{2}\right\rangle _{NI}}$. Under this transformation, with the use of Eqs.
\ref{lambdavsN} and \ref{kfvsN}, and in the limit where $N\rightarrow\infty$,
the hyperradial Schr\"{o}dinger equation becomes%
\end{subequations}
\begin{equation}
\left(  \dfrac{-1}{2m^{\ast}}\dfrac{d^{2}}{dR^{\prime2}}+\dfrac{V_{eff}\left(
R^{\prime}\right)  }{E_{NI}}-E^{\prime}\right)  R^{\prime\left(  3N-1\right)
/2}F\left(  R^{\prime}\right)  =0 \label{Heffrescale}%
\end{equation}
where $m^{\ast}=\left(  \lambda+3N/2\right)  ^{2}$. The effective potential
takes on the simple form
\begin{equation}
\dfrac{V_{eff}\left(  R^{\prime}\right)  }{E_{NI}}\rightarrow\dfrac
{1}{2R^{\prime2}}+\dfrac{1}{2}R^{\prime2}+\dfrac{\sigma k_{f}\gamma}%
{R^{\prime3}}. \label{sVeffrescale}%
\end{equation}
where $\sigma=1024/2835\pi^{3}$ and $\gamma=g/\hbar\omega l^{2}$. We note that
the only parameter that remains in this potential is the dimensionless
quantity $k_{f}\gamma$ and that in oscillator units $\gamma=g$.

In the limit where $N\rightarrow\infty$ the effective mass $m^{\ast}$ becomes%
\begin{equation}
m^{\ast}\rightarrow\dfrac{1}{16}\left(  3N\right)  ^{8/3}.
\label{masseffective}%
\end{equation}
For large numbers of particles, the second derivative terms in Eq.
\ref{Heffrescale} becomes negligible, a fact used later in Section IIIc.

\begin{figure}[h]
\begin{center}
\includegraphics[width=3.5in]{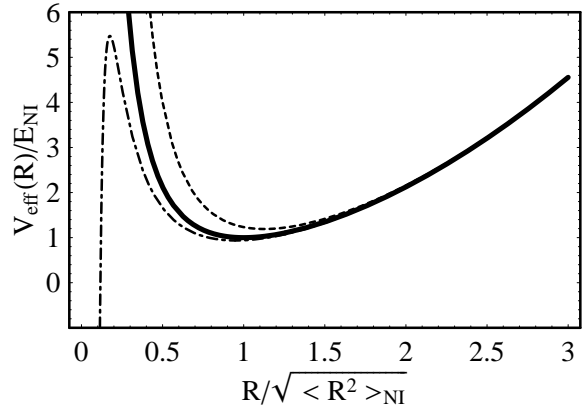}
\end{center}
\caption{The effective potential as a function of the hyperradius for
$k_{f}\gamma=0$ (solid), $k_{f}\gamma=15$ (dashed) and $k_{f}\gamma=-5.$
(dot-dash)}%
\label{Veffplot1}%
\end{figure}

The behavior of $V_{eff}\left(  R^{\prime}\right)  $ versus $R^{\prime}$ is
illustrated in Fig. \ref{Veffplot1} for various values of $k_{f}\gamma$. For $k_{f}\gamma=0$ (solid
curve), the non-interacting limit, the curve is exact and the ground state
solution is given by equation \ref{NIhyperrad}. For non-zero values of $g,$
$V_{eff}$ acquires an attractive ($k_{f}\gamma<0$) or repulsive ($k_{f}%
\gamma>0$) $1/R^{\prime3}$ contribution as indicated by the dot-dashed and
dashed lines respectively. For $k_{f}\gamma<0$ the DFG is metastable in a
region which has a repulsive barrier which it may tunnel through and emerges
in the region of small $R^{\prime}$ where the interaction term is dominant. It
should be noted, though, that small $R^{\prime}$ means the overall size of the
gas is small. Thus the region of collapse corresponds to a very high density
in the gas. In this region several of the assumptions made can fall apart,
most notably the assumption dealing with the validity of the two- body,
zero-range potential.\cite{esry1999vsi} For $k_{f}\gamma>0$ the positive
$1/R^{\prime3}$ serves to strengthens the repulsive barrier and pushes the gas
further out.

\subsection{Repulsive interactions ($g>0$)}

\begin{figure}[h]
\begin{center}
\includegraphics[width=3.5in]{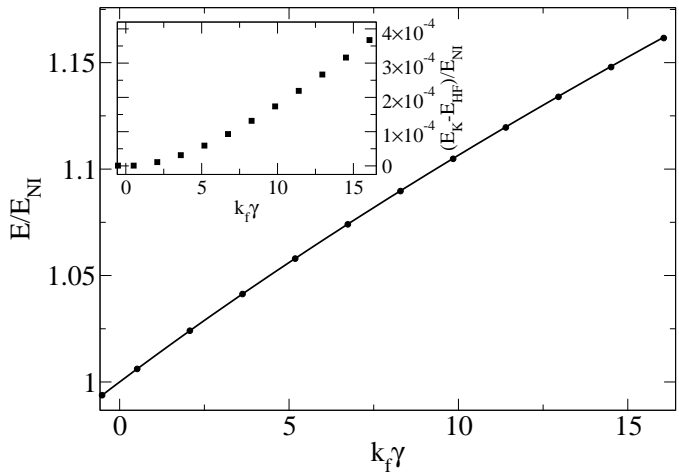}
\end{center}
\caption{The ground state energy in units of the non-interacting energy versus
$k_{f}\gamma$ for 240 atoms calculated using the K harmonic method (curve) and
using Hartree-Fock (circles). Inset: the difference in the ground state
energies predicted be the K harmonic ($E_{K}$) and Hartree-Fock ($E_{HF}$).
Clearly the K harmonic energies are slightly higher than Hartree-Fock.}%
\label{Erepulfig}%
\end{figure}

For positive values of the interaction parameter $g$ we expect the predicted
energy for this K harmonic method to deviate from experimental values, since
the trial wavefunction does not allow any fermions to combine into molecular
pairs as has been seen in
experiments.\cite{regal2004orc,zwierlein2004cpf,bartenstein2004cmb,bourdel2004esb}
Our method only can describe the normal degenerate fermi gas. The strong
repulsive barrier for repulsive interactions shown in Fig. \ref{Veffplot1}
arises as the gas pushes against itself which increases the energy and rms
radius of the ground state. Figs. \ref{Erepulfig} \begin{figure}[hh]
\begin{center}
\includegraphics[width=3.5in]{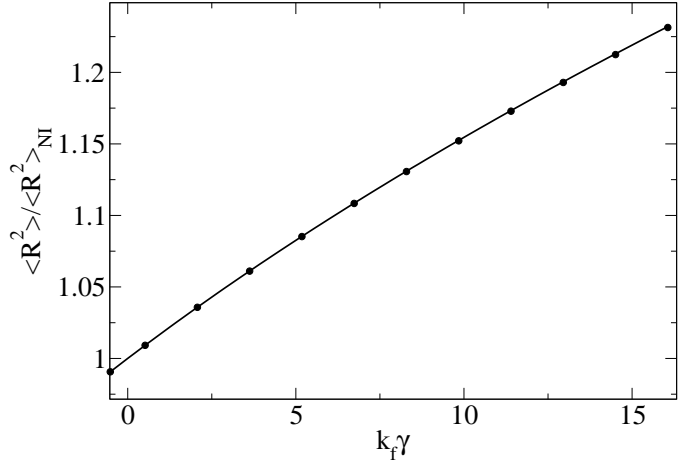}
\end{center}
\caption{The ground state average squared radius of the gas atoms in units of
the non-interacting rms squared radius is plotted versus $k_{f}\gamma$. The
calculations considered 240 atoms in both the K harmonic method (curve) and
Hartree-Fock (squares).}%
\label{Rrepulfig}%
\end{figure}and \ref{Rrepulfig} compare the ground state energy and average
radius squared respectively of 240 trapped atoms, plotted as a function of
$k_{f}\gamma$ with a Hartree-Fock (HF) calculation. The inset in Fig.
\ref{Erepulfig} shows that the K harmonic energies are slightly above the HF
energies; since both methods are variational upper bounds, we can conclude
that the Hartree-Fock solution is a slightly better representation of the true
solution to the full Schr\"{o}dinger equation with $\delta$-function interactions.

\begin{figure}[ptb]
\begin{center}
\includegraphics[width=3in]{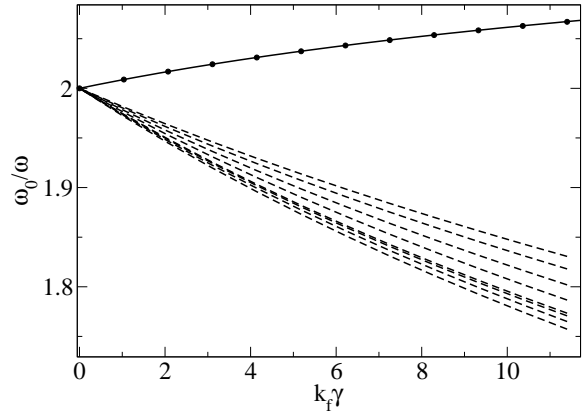}
\end{center}
\caption{The lowest breathing mode excitation ($\omega_{0}$) in units of the
trap frequency is plotted versus $k_{f}\gamma$ for the K harmonic method
(solid curve) and for the sum rule (circles). Also shown as dashed curves are
the lowest eight radial excitation frequencies predicted in the Hartree-Fock
approximation.}%
\label{freqrepuls}%
\end{figure}

An added benefit of the K harmonic method is that we now have an intuitively
simple way to understand the energy of the lowest radial excitation of the
gas, i.e. the breathing mode frequency. Fig. \ref{Veffplot1} shows that as
$k_{f}\gamma$ increases the repulsion increases the curvature at the local
minimum, whereby stronger repulsion causes the breathing mode frequency to
increase. Fig. \ref{freqrepuls} compares the breathing mode frequency calculated using the K
harmonic method to the sum rule prediction \cite{vichi1999coi} based on HF
orbitals, and also the lowest eight radial excitation frequencies predicted by
Hartree-Fock. As anticipated, the K harmonic method and the sum rule method
agree that the breathing mode frequency will increase with added repulsion.
Interestingly both the K harmonic and sum rule methods disagree qualitatively
with all eight of the lowest HF excitations. This difference is attributed to
the fact that Hartree-Fock on its own can only describe \emph{single particle
}excitations while both the sum rule and the K harmonic methods describe
\emph{collective }excitations in which the entire gas oscillates coherently.

As another test of the K harmonic method we calculate the peak density of the
gas. To do this we first define the density%
\begin{equation}
\rho\left(  \vec{r}\right)  =\int\left(  \prod\limits_{j=1}^{N}d^{3}%
r_{j}\right)  \sum_{i=1}^{N}\delta^{3}\left(  \vec{r}_{i}-\vec{r}\right)
\left\vert \Psi\right\vert ^{2}.\label{densedef}%
\end{equation}
It can be seen that integration over $\vec{r}$ using this definition gives
$\int\rho\left(  \vec{r}\right)  d^{3}r=N$. We recall that our separable
approximation takes the form $\Psi=F\left(  R\right)  \Phi_{\lambda}\left(
\Omega\right)  $. Use of \ref{densedef} and the antisymmetry of $\Phi
_{\lambda}$ gives%
\begin{equation}
\rho\left(  0\right)  =N\int dRR^{3N-1}\left\vert F\left(  R\right)
\right\vert ^{2}\int d\Omega\delta^{3}\left(  \vec{r}_{N}\right)  \left\vert
\Phi_{\lambda}\left(  \Omega\right)  \right\vert ^{2}.\label{dense1}%
\end{equation}
The hyperangular integration is carried out in Appendix D. The result is%
\begin{equation}
\dfrac{\rho\left(  0\right)  }{\rho_{NI}\left(  0\right)  }=\xi\int
dR\dfrac{R^{3N-1}\left\vert F\left(  R\right)  \right\vert ^{2}}{R^{3}%
}.\label{dens}%
\end{equation}
where $\rho_{NI}\left(  0\right)  $ is the non-interacting peak density and
$\xi=\dfrac{l^{3}\Gamma\left(  \lambda+3N/2\right)  }{N^{3/2}\Gamma\left(
\lambda+3\left(  N-1\right)  /2\right)  }$. For the $n$th filled energy shell
the non-interacting peak density is given by%
\[
\rho_{NI}\left(  0\right)  \approx\dfrac{k_{f}^{3}}{6\pi^{2}}=\dfrac{1}%
{6\pi^{2}}\left[  \dfrac{2m\omega}{\hbar}\left(  n+1/2\right)  \right]  ^{3/2}%
\]
Note that the peak density is \emph{not} given by $R^{3N-1}\left\vert F\left(
R\right)  \right\vert ^{2}$ evaluated at $R=0$, this describes the probability
of all of the particles being at the center at once.

\begin{figure}[ptb]
\begin{center}
\includegraphics[width=3in]{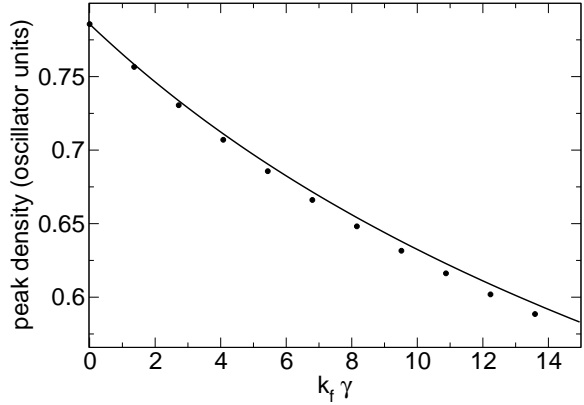}
\end{center}
\caption{The peak density in units of $\left(  \hbar/m\omega\right)  ^{-3/2}$
versus $k_{f}\gamma$ predicted by the K harmonic (solid) and HF (circles).
Both sets of calculations were done for a filled shell of 240 atoms.}%
\label{rhorepulsfig}%
\end{figure}

Fig. \ref{rhorepulsfig} compares the K harmonic and HF peak densities. Clearly the density
does decrease from the non-interacting value. The two methods are in good
qualitative agreement, but Hartree-Fock seems to predict a slightly lower
density, presumably a manifestation of the slightly inferior K harmonic
wavefunction, $F\left(  R\right)  \Phi_{\lambda}\left(  \Omega\right)  $.

\subsection{Attractive interactions ($g<0$)}

\begin{figure}[ptb]
\begin{center}
\includegraphics[width=3in]{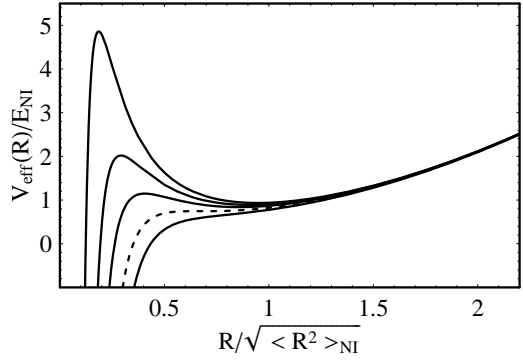}
\end{center}
\caption{$V_{eff}$ versus $R^{\prime}$ for several values of $k_{f}\gamma$.
$k_{f}\gamma=k_{f}\gamma_{c}$ (dashed) and from top to bottom $k_{f}%
\gamma=-5.3,-8.3,-11.3,-19.3$ (all solid)}%
\label{Veffplot2}%
\end{figure}

In this section we examine the behavior of the gas under the influence of
attractive s-wave interactions ($k_{f}\gamma<0$) . For attractive interactions
the gas lives in a metastable region and can tunnel through the barrier shown
in Fig. \ref{Veffplot1}. Fig. \ref{Veffplot2} shows the behavior of $V_{eff}$ for several values of $k_{f}%
\gamma$. The location of the local minimum gets pulled down with stronger
attraction as the gas pulls in on itself and deeper into the center of the
trap. Further, as the strength of the interaction increases, the height of the
barrier decreases. In fact beyond a critical interaction strength $\gamma_{c}$
the interaction becomes so strong that it always dominates over the repulsive
kinetic term. At this critical point the local extrema disappear entirely and
the gas is free to fall into the inner \textquotedblleft collapse" region. The
value of $\gamma_{c}$ can be calculated approximately by finding the point
where $V_{eff}$ loses its local minimum and becomes entirely attractive. This
is not exact as the gas will have some small zero point energy that will allow
it to tunnel through or spill over the barrier before the minimum entirely
disappears. This critical interaction strength is given by%
\[
k_{f}\gamma_{c}=-\dfrac{189\pi^{3}}{256}\dfrac{1}{5^{1/4}}\approx-15.31.
\]
\qquad

Just before the minimum disappears, its location is given by $R_{\min}%
^{\prime}=5^{-1/4}$, with an energy of $V_{eff}\left(  R_{\min}^{\prime
}\right)  =\sqrt{5}E_{NI}/3\approx0.75E_{NI}$. This means that if the gas is
mechanically stable for all values of the two body scattering length, i.e.
$a\rightarrow-\infty$, in this approximation, there must be a renormalization
cutoff in the strength of the $\delta$-function such that $k_{f}\gamma>-15.31$
for all $a$. With this in mind, we begin to examine the behavior of the DFG
for the allowed values of $k_{f}\gamma$.

\begin{figure}[ptb]
\begin{center}
\includegraphics[width=3in]{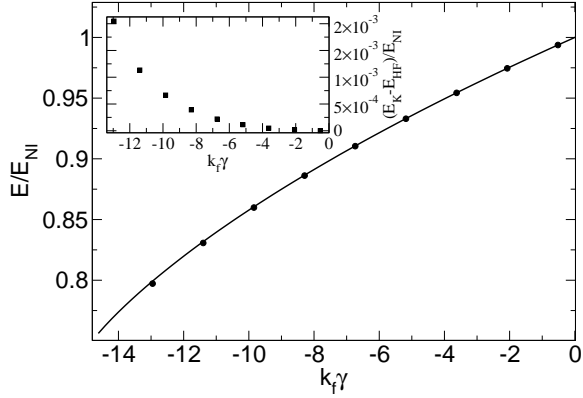}
\end{center}
\caption{The ground state energy (in units of the non-interacting energy)
versus $k_{f}\gamma$ for 240 atoms calculated using the K harmonic (curve) and
Hartree-Fock (circles) methods. Inset: The difference in the ground state
energies predicted be the K harmonic ($E_{K}$) and Hartree-Fock ($E_{HF}$).
Clearly the K harmonic prediction is slightly higher.}%
\label{Eattracfig}%
\end{figure}

\begin{figure}[ptbptb]
\begin{center}
\includegraphics[width=3in]{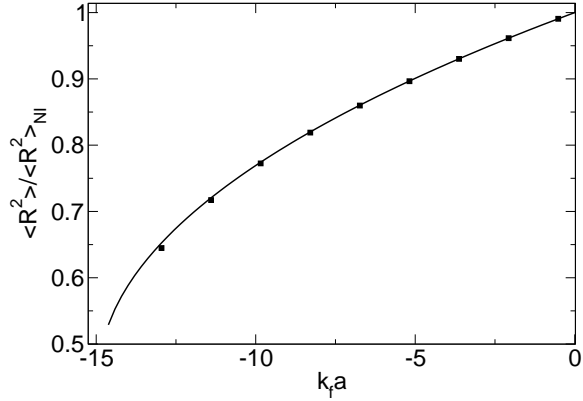}
\end{center}
\caption{The average squared radius of the Fermi gas ground state in units of
the non-interacting value is plotted versus $k_{f}\gamma$. The calculations
are for 240 atoms in both the K harmonic method (curve) and Hartree-Fock
(squares).}%
\label{Rattracfig}%
\end{figure}

\begin{figure}[ptbptbptb]
\begin{center}
\includegraphics[width=3in]{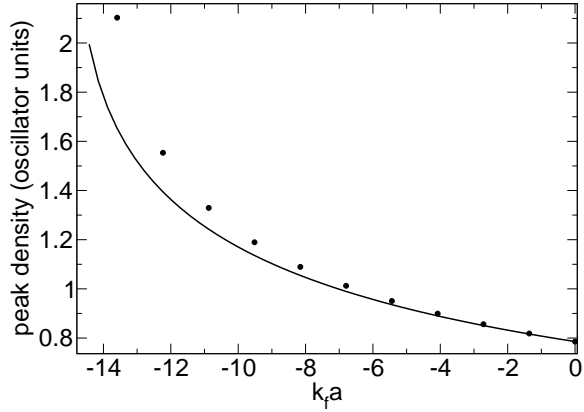}
\end{center}
\caption{The peak density in units of $\left(  \hbar/m\omega\right)  ^{-3/2}$
versus $k_{f}\gamma$ predicted by the K harmonic (solid line) and HF
(circles). Both calculations were carried out for a filled shell of 240
atoms.}%
\label{rhoattrac}%
\end{figure}

Figs. \ref{Eattracfig} and \ref{Rattracfig} show a comparison of the ground
state energy and rms radius of the gas versus $k_{f}\gamma$ down to the
$k_{f}\gamma_{c}$ as calculated in the K harmonic and Hartree-Fock methods.
Again Hartree-Fock does just slightly better in energy, which we interpret as
Hartree-Fock giving a slightly better representation of the actual ground
state wavefunction. The energy difference becomes largest as the interaction
strength approaches the critical value. This increase is due to the fact that
Hartree-Fock predicts that collapse occurs slightly earlier with $k_{f}%
\gamma_{c}\approx-14.1$. As the interaction strength increases, the energy and rms radius
of the gas decrease. As $k_{f}\gamma$ approaches $k_{f}\gamma_{c}$ the overall
size of the gas decreases sharply and from Eq. \ref{dens} we expect to see a
very sharp rise in the peak density. This behavior is apparent in Fig.
\ref{rhoattrac}, which displays the peak density versus $k_{f}\gamma$ for both
the K harmonic and Hartree-Fock.

\begin{figure}[ptb]
\begin{center}
\includegraphics[width=3in]{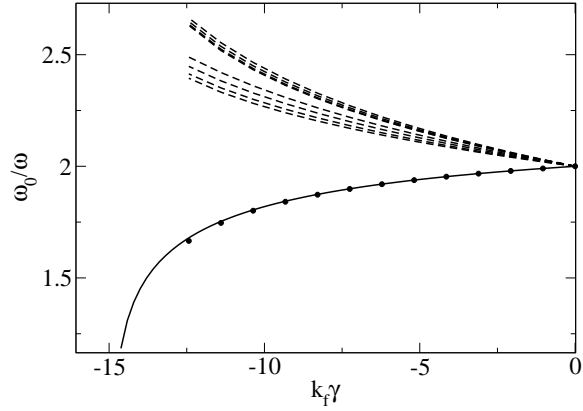}
\end{center}
\caption{The frequency of the lowest energy radial transition in units of the
trap frequency versus $k_{f}\gamma$ predicted by the K harmonic method (solid
line) and by the sum rule (circles). Also shown are the lowest eight radial
transitions predicted by Hartree-Fock.}%
\label{freqattrac}%
\end{figure}

\begin{figure}[ptbptb]
\begin{center}
\includegraphics[width=3in]{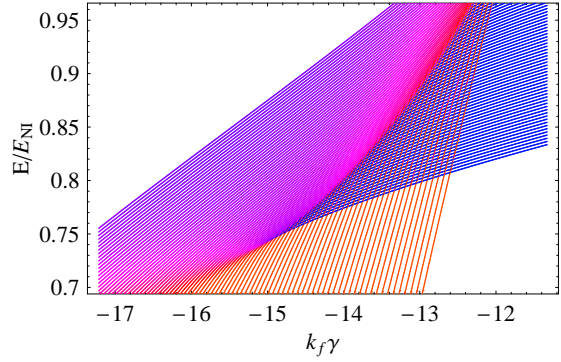}
\end{center}
\caption{(color online) A portion of the energy spectrum vs $k_{f}\gamma$
close to the critical point $k_{f}\gamma=k_{f}\gamma_{c}$. Levels in the
metastable region (blue) decrease slowly while levels in the collapse region
(red) decrease very quickly. Energy levels above the barrier in $V_{eff}^{s}$
(purple) live both in the collapse region and the metastable region.}%
\label{Enlevs}%
\end{figure}

While the local minimum present in $V_{eff}$ only supports metastable states,
it is still informative to examine the behavior of the energy spectrum versus
$k_{f}\gamma$, beginning with the breathing mode frequency. As the interaction
strength becomes more negative Fig. \ref{Veffplot2} shows that the curvature
about the local minimum in $V_{eff}$ decreases. This \textquotedblleft
softening" of the hyperradial potential leads to a decrease in the breathing
mode frequency in the outer well. Fig. \ref{freqattrac} shows the breathing
mode vs. $k_{f}\gamma$ predicted by the K harmonic (curve) method and also
using the sum rule with Hartree-Fock orbitals (circles). Also shown in Fig.
\ref{freqattrac} are the lowest eight Hartree-Fock excitation frequencies for
a filled shell of 240 atoms. Again, the K harmonic method agrees quite well
with the sum rule, while both differ qualitatively from the HF prediction. The
sharp decrease in the breathing mode frequency that occurs as $k_{f}%
\gamma\rightarrow k_{f}\gamma_{c}$ is a result of the excited mode
\textquotedblleft falling" over the barrier into the collapse region as the
barrier is pulled down by the interaction. Figure \ref{Enlevs} displays some energy levels in the metastable
region as functions of $k_{f}\gamma$ near $k_{f}\gamma_{c}$. Because of the
singular nature of the $1/R^{3}$ behavior in the inner region we have added an
inner repulsive $1/R^{12}$ barrier to truncate the infinitely many nodes of
the wavefunction in the inner region. The behavior of the wavefunction is not
correct within this region anyway because recombination would become dominant,
and in any case the zero-range interaction is suspect beyond $\left\vert
k_{f}a\right\vert \sim1$ and it must be renormalized. Figure \ref{Enlevs} shows three distinct types of energy level.
Levels that are contained in the local minimum (shown in blue) are decreasing,
but not as quickly as the others; levels that are in the collapse region
(shown in red) have a very steep slope as they are drawn further into toward
$R=0$; and energy levels that are above the barrier in $V_{eff}$ (shown in
purple) have wavefunctions are in both the collapse region and the local
minimum. As $k_{f}\gamma$ decreases, the higher energy levels fall over the
barrier into the collapse region earlier, until finally just before
$k_{f}\gamma_{c}$ is reached the lowest metastable level falls below the
\textquotedblleft ground state". This corresponds to the breathing mode
behavior seen in Fig. \ref{freqattrac}. Of course all of this applies only if
there is no further hyperradial dependence of $g$. If the interaction becomes
density dependent, as is the case in some types of renormalization, a change
in the hyperradius will change the density and thus the interaction coupling
parameter, i.e. $g\rightarrow g\left(  R\right)  $.

\subsection{The large $N$ limit}

In Sections IIIa and IIIb the ground state energy and expectation values
discussed were found by solving the hyperradial effective Schr\"{o}dinger
equation \ref{HypersRH} for a finite number of particles. Here we discuss the
behavior of the $N$ fermion system in the limit where $N$ is large. To do this
we exploit the fact that the $\partial^{2}/\partial R^{2}$ term in the
effective Hamiltonian, the \textquotedblleft hyperradial kinetic energy",
becomes negligible. In this limit we see that the total energy of the system
is merely given by $E=V_{eff}\left(  R_{\min}\right)  $, as is the case in
dimensional perturbation theory.\cite{Germann95} To find the ground state
energy we must merely find the minimum (local minimum for $g<0$) value of
$V_{eff}$. Accordingly, we find the roots of $\dfrac{dV_{eff}}{dR^{\prime}}%
=0$. Using Eq. \ref{sVeffrescale} we simplify this to%
\begin{equation}
k_{f}\gamma=\dfrac{1}{3\sigma}R_{\min}^{\prime}\left(  R_{\min}^{\prime
4}-1\right)  \label{mineq}%
\end{equation}
\begin{figure}[ptb]
\begin{center}
\includegraphics[width=3in]{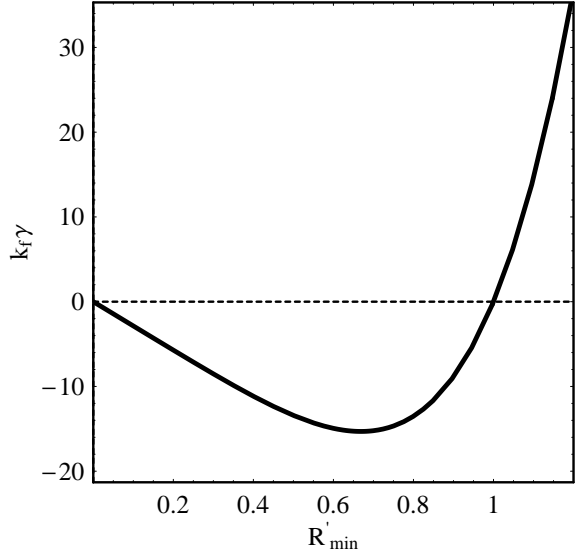}
\end{center}
\caption{Plot of the potential curve extrema which obey (\ref{mineq}).
Examination of this plot tells us the behavior of $R_{\min}^{\prime}$ for all
allowed values of $k_{f}\gamma$ including the existence of the critical point
$k_{f}\gamma_{c}$ located at the minimum of the plot where the maximum and
minimum coincide.}%
\label{minplot}%
\end{figure}where $R_{\min}^{\prime}$ is the hyperradial value that minimizes
$V_{eff}$. The solutions to Eq. \ref{mineq} are illustrated graphically in
Fig. \ref{minplot}; for any given $k_{f}\gamma$ we need only look for the
value of $R_{\min}^{\prime}$ that gives that value. The exact solution of Eq.
\ref{mineq} cannot be determined analytically for all values of $k_{f}\gamma,$
but Fig. \ref{minplot} shows that for $k_{f}\gamma>0$ there is always only one
positive, real $R_{\min}^{\prime}$ that satisfies Eq. \ref{mineq}. This
corresponds to the global minimum discussed for repulsive interactions. For
$k_{f}\gamma<0$ things are a bit more complicated. Fig. \ref{minplot} shows
that for $k_{f}\gamma_{c}<k_{f}\gamma<0$ there are two solutions to
\ref{mineq}. The inner solution is a local maximum and corresponds to the peak
of the barrier seen in Fig. \ref{Veffplot2}$,$ the outer solution corresponds
to the local minimum where the DFG lives. The local minimum is the state that
we are concerned with here as this will give the energy and hyperradial
expectation values of the metastable Fermi gas. The value of $k_{f}\gamma$
where these two branches merge is the place where the local maximum merges
with the local minimum namely the critical value $k_{f}\gamma_{c}$. Any value
of $k_{f}\gamma$ less than $k_{f}\gamma_{c}$ has no solution to \ref{mineq}
and thus we cannot say that there is a region of stability. For $k_{f}%
\gamma=0$, the non-interacting limit, we see that there are two solutions
$R_{\min}^{\prime}=0$ and $R_{\min}^{\prime}=1$. The solution $R_{\min
}^{\prime}=0$ must be discounted as there is a singularity in $V_{eff}$ at
$R=0$. Thus in the non-interacting limit $R_{\min}^{\prime}\rightarrow1$ and
$V_{eff}\left(  R_{\min}\right)  \rightarrow E_{NI}$, as expected.

\begin{figure}[ptb]
\begin{center}
\includegraphics[width=3in]{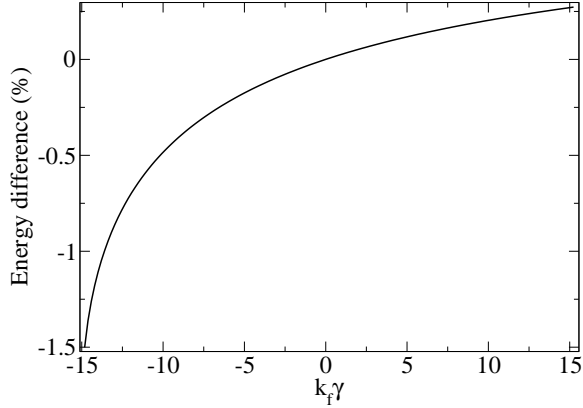}
\end{center}
\caption{The percentage difference between the energy found by minimizing
$V_{eff}$ and the energy found by explicitly solving the hyperradial
Schr\"{o}dinger equation for 240 atoms.}%
\label{Endifffig}%
\end{figure}

Substitution of \ref{mineq} into \ref{sVeffrescale} gives the energy of the
ground state in the large $N$ limit, as a function of the size of the gas.%
\[
\dfrac{V_{eff}\left(  R_{\min}\right)  }{E_{NI}}=\dfrac{1+5R_{\min}^{\prime4}%
}{6R_{\min}^{\prime2}}.
\]
These solutions to Eq. \ref{mineq} immediately give the ground state energy of
the gas versus $k_{f}\gamma.$ Fig. \ref{Endifffig} shows the percentage
difference of the ground state energy found by this minimization procedure and
that of 240 particles found by diagonalizing $H_{eff}$ in Eq. \ref{HypersRH}.
We see in Fig. \ref{Endifffig} that for $\left\vert k_{f}%
\gamma\right\vert <10$ the energy difference is less than 0.5\%. As
$\left\vert k_{f}\gamma\right\vert $ gets larger the difference increases, but
in the range shown the difference is always less than $\pm1.5\%$.

Another result from \ref{Heffrescale} is the fact that in the large $N$ limit
the commutator $\left[  H_{eff},R\right]  \rightarrow0$. Thus for any operator
that is solely a function of the hyperradius $\hat{O}\left(  R\right)  $ the
ground state expectation value in the large $N$ limit is given by the operator
evaluated at $R_{\min}^{\prime}$, i.e. $\left\langle \hat{O}\left(  R^{\prime
}\right)  \right\rangle =\hat{O}\left(  R_{\min}^{\prime}\right)  $. This
tells us that the large $N$ limit wavefunction is given by $\left[  R^{\left(
3N-1\right)  /2}G\left(  R\right)  \right]  ^{2}=\delta\left(  R-R_{\min
}\right)  $. We can perturb this slightly and say that the ground state
hyperradial wavefunction can be approximated by a very narrow Gaussian
centered at $R_{\min}$. To find the width of this gaussian we approximate
$V_{eff}$ about $R_{\min}$ as a harmonic oscillator with mass $m^{\ast}$ and
frequency $\omega_{0}$. We may find $\omega_{0}$ by comparing the oscillator
potential with the second order Taylor series about $R_{\min}$ in $V_{eff}$ ,
i.e.%
\begin{equation}
\omega_{0}=\sqrt{\dfrac{1}{m^{\ast}}\dfrac{1}{E_{NI}}\left(  \left.
\dfrac{\partial^{2}V_{eff}}{\partial R^{\prime2}}\right\vert _{R^{\prime
}=R_{\min}^{\prime}}\right)  }. \label{freqNIunit}%
\end{equation}
The effective hyperradial oscillator length of $R^{\prime\left(  3N-1\right)
/2}G\left(  R\right)  $ is then given by $l_{0}=\sqrt{\hbar/m^{\ast}\omega
_{0}}$.

The breathing mode frequency is now simply found to be $\omega_{0}.$ The
frequency in Eq. \ref{freqNIunit} is in units of the non-interacting energy,
so to get back to conventional units we multiply by $E_{NI}/\hbar$. From Eq.
\ref{Heffrescale} we know that $m^{\ast}=mE_{NI}N\left\langle R^{2}%
\right\rangle _{NI}/\hbar^{2}$. Noting that $N\left\langle R^{2}\right\rangle
_{NI}=l^{2}E_{NI}/\hbar\omega$ this leads to%
\begin{equation}
\omega_{0}^{B}=\sqrt{\dfrac{1}{E_{NI}}\left(  \left.  \dfrac{\partial
^{2}V_{eff}}{\partial R^{\prime2}}\right\vert _{R^{\prime}=R_{\min}^{\prime}%
}\right)  }. \label{ezbreath}%
\end{equation}
for $\omega_{0}^{B}$, the breathing mode frequency in units of the trap
frequency. Using Eq. \ref{sVeffrescale} and substituting in Eq. \ref{mineq} to
evaluate at the minimum gives that%
\[
\omega_{0}^{B}=\sqrt{5-\dfrac{1}{R_{\min}^{\prime4}}}.
\]
We note that this is now dependent only on the value of $k_{f}\gamma$, i.e.
for a fixed $k_{f}\gamma$ the predicted breathing mode is independent of the
number of atoms in the system in the large $N$ limit.

\begin{figure}[ptb]
\begin{center}
\includegraphics[width=3in]{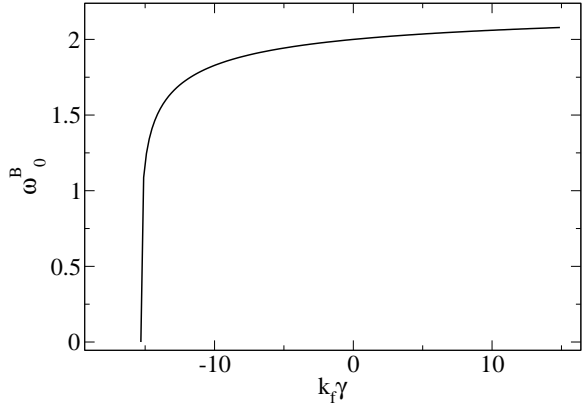}
\end{center}
\caption{The breathing mode (in oscillator units) in the large $N$ limit
versus $k_{f}\gamma$. Note that as $k_{f}\gamma\rightarrow k_{f}\gamma_{c}$
the frequency drops to zero as the local minimum disappears.}%
\label{BigNbreath}%
\end{figure}

This prediction can be compared with that predicted by the sum rule using the
formula found by the authors of Ref. \cite{vichi1999coi}:
\begin{equation}
\omega_{0}^{B}=\sqrt{4+\dfrac{3}{2}\dfrac{E_{int}}{E_{ho}}}.
\label{breathRmin}%
\end{equation}
Here $E_{int}$ and $E_{ho}$ are the expectation values of the interaction
potential and the oscillator potential in the ground state respectively. If we
insert the expectation values predicted here by the K harmonic method we find%
\[
\omega_{0}^{B}=\sqrt{4+3\dfrac{\sigma k_{f}\gamma}{R_{\min}^{\prime5}}}.
\]
Substituting \ref{mineq} for $k_{f}\gamma$ gives%
\[
\omega_{0}^{B}=\sqrt{5-\dfrac{1}{R_{\min}^{\prime4}}}%
\]
which agrees exactly with the frequency predicted in Eq. \ref{breathRmin}.
Fig. \ref{BigNbreath} shows the breathing mode frequency predicted by Eq.
\ref{breathRmin} versus $k_{f}\gamma$. We see the same behavior in this plot as was seen in Fig.
\ref{freqattrac} where the breathing mode frequency dives to zero as
$k_{f}\gamma\rightarrow k_{f}\gamma_{c}$.

\section{Summary and Prospects}

We have demonstrated an alternative approach to describing the physics of a
trapped degenerate Fermi gas from the point of view of an ordinary linear
Schr\"{o}dinger equation with two body, microscopic interactions. The use of a
hyperspherical variational trial wave function whose hyperangular behavior is
frozen to be that of the K harmonic yields a one-dimensional effective
potential (\ref{swavepot}) in a collective coordinate, the hyperradius $R$.
This approach yields an intuitive understanding of the energy and size of the
DFG in terms of familiar Schr\"{o}dinger quantum mechanics. Perhaps
surprisingly, this approximation gives good results for the ground state
energy, rms radius and peak number density of the gas. These are all in
quantitative agreement with HF, while the breathing mode frequency compares
well with that found using the sum rule method, but gives a qualitative
improvement over the lowest calculated HF radial excitation frequencies.

The work presented here has been limited to the case of filled energy shells,
i.e. \textquotedblleft magic numbers" of atoms, and it should be readably
generalizable, with minor extensions, to open shells as well. The present
results are limited to a spherically symmetric trap. Generalization to an
cylindrically symmetric \textquotedblleft cigar" trap should be possible with
a judicious choice of coordinates, and will be presented elsewhere.

For strongly attractive interactions, this picture predicts an instability in
the DFG\ in a manner qualitatively similar to the physics of the
\textquotedblleft Bosenova".\cite{bohn_esry_greene_hsbec} For the gas to
remain stable across the BEC-BCS crossover regime the strength of interaction
potential must be bounded from below, $k_{f}\gamma>-15.31$. Preliminary
results from another study \cite{Fregoso06} indicate that renormalization of
the singular $\delta$-function interaction accomplishes precisely that, and
apparently prevents collapse. The full interrelation between this picture and
that of pairing in the BEC-BCS crossover region is beyond the scope of this
study and is a subject that will be relegated to future publications.

\section{ACKNOWLEDGMENTS}

\qquad This work was supported by funding from the National Science
Foundation. We thank N. Barnea for introducing us to refs.
\cite{Timofeyuk04,Timofeyuk02}. Discussions with L. Radzihovsky and V. Gurarie
have also been informative.

\section*{APPENDIX A: HYPERRADIAL SOLUTION TO THE NON-INTERACTING OSCILLATOR}

We wish to derive Eq. \ref{NIhyperrad}, the nodeless hyperradial solution to
the Schr\"{o}dinger equation for $N$ non-interacting particles in an isotropic
oscillator. The Schr\"{o}dinger equation for this system is given by%
\begin{equation}
\left[  \sum_{i=1}^{N}\left(  \dfrac{-\hbar^{2}}{2m}\nabla_{i}^{2}+\dfrac
{1}{2}m\omega^{2}r_{i}^{2}\right)  -E\right]  \Psi=0 \tag{A1}\label{noninth}%
\end{equation}
where $\vec{r}_{1},\vec{r}_{2},...,\vec{r}_{N}$ are the cartesian coordinates
for each atom from the trap center.

To begin we examine the radial Schr\"{o}dinger equation for a single particle
in an isotropic trap%
\begin{equation}
\left(  \dfrac{-\hbar^{2}}{2m}\left(  \dfrac{d^{2}}{dr^{2}}-\dfrac{\ell\left(
\ell+1\right)  }{r^{2}}\right)  +\dfrac{1}{2}m\omega^{2}r^{2}-E_{n\ell
}\right)  rf_{n\ell}\left(  r\right)  =0. \tag{A2}\label{singleNI}%
\end{equation}
The solution to this is well known and is given by%
\begin{equation}
rf_{n\ell}\left(  r\right)  =A_{n\ell}\exp\left(  -r^{2}/2l\right)  \left(
\dfrac{r}{l}\right)  ^{\ell+1}L_{n}^{\ell+1/2}\left(  \dfrac{r^{2}}{l^{2}%
}\right)  \tag{A3}\label{indepsol}%
\end{equation}
with energy $E_{n\ell}=\hbar\omega\left(  2n+\ell+3/2\right)  $ where
$l=\sqrt{\hbar/m\omega}$ and $n$ is the number of radial nodes in the wavefunction.

Transformation of (\ref{noninth}) into hyperspherical coordinates using Eqs.
\ref{hyperr}, \ref{hypera}, \ref{kineticE} and \ref{oscillatorpot} yields a
Schr\"{o}dinger equation that separates into hyperradial and hyperangular
pieces. The hyperangular solution is a hyperspherical harmonic that
diagonalizes $\mathbf{\Lambda}^{2}.$ The resulting hyperradial Schr\"{o}dinger
equation is given by%
\begin{equation}
\left(  \dfrac{-\hbar^{2}}{2R}\left(  \dfrac{d^{2}}{dR^{2}}-\dfrac{K\left(
K+1\right)  }{R^{2}}\right)  +\dfrac{1}{2}M\omega^{2}R^{2}-E\right)
R^{\left(  3N-1\right)  /2}F\left(  R\right)  =0. \tag{A4}\label{HyperHNI}%
\end{equation}
where $K=\lambda+3\left(  N-1\right)  /2$. Comparing this with Eq.
\ref{singleNI} we see that if we make the substitutions $\ell\rightarrow K$,
$n\rightarrow\chi$, $m\rightarrow M$, $r\rightarrow R$ and $rf_{n\ell}\left(
r\right)  \rightarrow R^{\left(  3N-1\right)  /2}F_{\chi K}\left(  R\right)  $
the single particle radial Schr\"{o}dinger equation becomes the $N$ particle
hyperradial Schr\"{o}dinger equation. With these replacements the solution is
evidently%
\begin{equation}
R^{\left(  3N-1\right)  /2}G_{\chi K}\left(  R\right)  =A_{\chi K}\exp\left(
-R^{2}/2\mathcal{L}\right)  \left(  \dfrac{R}{\mathcal{L}}\right)
^{K+1}L_{\chi}^{K+1/2}\left(  \dfrac{R^{2}}{\mathcal{L}^{2}}\right)  ,
\tag{A5(a)}\label{HSsolution}%
\end{equation}
where $\mathcal{L}=l/\sqrt{N}$ and $\chi$ is the number of hyperradial nodes
in the $N$ body system. For $\chi=0$ this is the same hyperradial wavefunction
written in Eq. \ref{NIhyperrad}. The total energy is given by%
\begin{equation}
E_{\chi K}=\hbar\omega\left(  2\chi+K+\dfrac{3}{2}\right)  =\hbar\omega\left(
2\chi+\lambda+\dfrac{3N}{2}\right)  \tag{A5(b)}\label{HSenergy}%
\end{equation}
Now that we have a hyperspherical solution we can compare it with the solution
to Eq. \ref{noninth} written in terms of independent particle coordinates
$\left(  \vec{r}_{1},\vec{r}_{2},...,\vec{r}_{N}\right)  $. This equation is
clearly separable in each coordinate $\vec{r}_{i}$, and its solution is a
product of $N$ single particle wavefunctions
\[
\Psi=\prod\limits_{i=1}^{N}f_{n_{i}\ell_{i}}\left(  r\right)  y_{\ell_{i}%
m_{i}}\left(  \omega_{i}\right)
\]
Where $f_{n_{i}\ell_{i}}\left(  r\right)  $ is given by Eq. \ref{indepsol},
$\omega_{i}$ are the spherical polar angular coordinates of the $i$th particle
and $y_{\ell m}\left(  \omega\right)  $ is a normal 3D spherical harmonic. The
energy for this independent particle solution is given by%
\begin{equation}
E=\hbar\omega\left[  \dfrac{3N}{2}+\sum_{i=1}^{N}\left(  2n_{i}+\ell
_{i}\right)  \right]  . \tag{A6}\label{IndEnergy}%
\end{equation}
Now that we have seen that Eq. \ref{noninth} is separable in both
hyperspherical and independent particle coordinates, we may compare Eqs.
\ref{HSenergy} and \ref{IndEnergy} to find that%
\begin{equation}
2\chi+\lambda=\sum_{i=1}^{N}\left(  2n_{i}+\ell_{i}\right)  . \tag{A7}%
\label{lambrelation}%
\end{equation}
We should note that this does not mean that the independent particle solution
and the hyperspherical solution are the exact same, only that the
hyperspherical solution $F\left(  R\right)  \Phi_{\lambda}\left(
\Omega\right)  $ must be a linear combination of independent particle
solutions of the same energy.

\section*{APPENDIX B: PROOF OF THEOREM \ref{theorem1}}

Here we will use the results from Appendix A to prove theorem \ref{theorem1}.
We proceed by assuming that there exists a fully antisymmetric, ground state,
hyperspherical solution to the Schr\"{o}dinger equation for $N$
non-interacting fermions in an isotropic trap with energy given by Eq.
\ref{HSenergy} with $\chi>0$. If we can show that this leads to a
contradiction then we have, from equation \ref{lambrelation}, that there is
only one $\lambda$ for all of the ground state configurations. From Appendix A
we know that this hyperspherical solution must be a linear combination of
antisymmetric, degenerate, ground state solutions in independent particle
coordinates%
\[
F_{\chi K}\left(  R\right)  \Phi_{\lambda}\left(  \Omega,\sigma_{1},\sigma
_{2},...\sigma_{N}\right)  =\sum_{\nu}D_{\nu}\left(  \vec{r}_{1},\vec{r}%
_{2},...,\vec{r}_{N},\sigma_{1},\sigma_{2},...\sigma_{N}\right)
\]
where each $D_{\nu}$ is of the form of Eq. \ref{Slater1}, $\nu$ is used to
distinguish between degenerate states of the same energy and $F_{\chi
K}\left(  R\right)  $ is given by Eq. \ref{HSsolution}. Again, for brevity, we
suppress the spin coordinates $\left(  \sigma_{1},\sigma_{2},...\sigma
_{N}\right)  $ in $\Phi_{\lambda}$. We now note that the hyperradius $R$ is
completely symmetric under all transpositions of particle coordinates, thus
all of the transposition symmetry must be contained in the function
$\Phi_{\lambda}\left(  \Omega\right)  $. From $\Phi_{\lambda}\left(
\Omega\right)  $ we construct a new completely antisymmetrized hyperspherical
solution $F_{0K}\left(  R\right)  \Phi_{\lambda}\left(  \Omega\right)  $ which
has energy
\[
E=\hbar\omega\left(  \lambda+\dfrac{3N}{2}\right)  .
\]
Use of Eqs. \ref{HSenergy} and \ref{lambrelation} gives
\begin{equation}
\lambda=\dfrac{E_{NI}}{\hbar\omega}-\dfrac{3N}{2}-2\chi, \tag{B1}%
\label{lambdamin}%
\end{equation}
where $E_{NI}=\hbar\omega\left[  3N/2+\sum_{i=1}^{N}\left(  2n_{i}+\ell
_{i}\right)  \right]  $ is the ground state energy as defined by any of the
functions $D_{\nu}\left(  \left\{  \vec{r}_{i}\right\}  _{i=1}^{N}\right)  $.
Thus our new function $F_{0K}\left(  R\right)  \Phi_{\lambda}\left(
\Omega\right)  $ has energy%
\[
E=E_{NI}-2\chi
\]
which would lie below the ground state energy, a contradiction unless $\chi=0$.

In the above analysis we assumed a degenerate set of solutions at the ground
state energy. For non-degenerate solutions the proof becomes trivial, as any
nondegenerate solution must be the same no matter what coordinate system it is
expressed in. From this the rest of the proof follows in the same way, and the
final, unique $\lambda$ for this system is given by Eq. \ref{lambdamin},
\begin{equation}
\lambda=\dfrac{E_{NI}}{\hbar\omega}-\dfrac{3N}{2}. \tag{B2}\label{lambtot}%
\end{equation}

\section*{APPENDIX C: CALCULATION OF THE S-WAVE INTERACTION MATRIX ELEMENT}

To calculate $C_{N}$ it is useful to start with a more general interaction. We
assume that the interaction term in the total $N$ body Hamiltonian is such
that at a fixed hyperradius $U_{int}\left(  \vec{r}_{ij}\right)  $ is
separable into a hyperradial function times a hyperangular integral, i.e.%
\begin{equation}
U_{int}\left(  \vec{r}_{ij}\right)  \equiv V_{ij_{R}}\left(  R\right)
V_{ij_{\Omega}}\left(  \Omega\right)  .\tag{C1}\label{factoredint}%
\end{equation}
From properties of the $\delta$-function and Eq. \ref{hypera2} it is easy to
se that the $U_{int}\left(  \vec{r}\right)  =g\delta^{3}\left(  \vec
{r}\right)  $ fits this criteria. While $V_{ij_{\Omega}}\left(  \Omega\right)
$ might have some very complex form, it will be seen shortly that only the
form of $V_{ij_{R}}\left(  R\right)  $ and $U_{int}\left(  \vec{r}%
_{ij}\right)  $ will matter.
\[
U_{eff}\left(  R\right)  =\left\langle \Phi_{\lambda}\left\vert \sum
_{i>j}U_{int}\left(  \vec{r}_{ij}\right)  \right\vert \Phi_{\lambda
}\right\rangle
\]
we may again use the antisymmetry of $\Phi_{\lambda}$ to exchange $\vec{r}%
_{i}\rightarrow\vec{r}_{2}$ and $\vec{r}_{j}\rightarrow\vec{r}_{1}$ and arrive
at%
\[
U_{eff}\left(  R\right)  =\dfrac{N\left(  N-1\right)  }{2}\left\langle
\Phi_{\lambda}\left\vert U_{int}\left(  \vec{r}_{21}\right)  \right\vert
\Phi_{\lambda}\right\rangle .
\]
From Eq. \ref{factoredint} we can write that%
\begin{equation}
U_{eff}\left(  R\right)  =\zeta V_{R}\left(  R\right)  \tag{C2}\label{Ueffgen}%
\end{equation}
Where $\zeta=\left\langle \Phi_{\lambda}\left\vert V_{\Omega}\left(
\Omega\right)  \right\vert \Phi_{\lambda}\right\rangle $ is the analogue of
$gC_{N}/N^{3/2}$ in Eq. \ref{Ueff}. To find $\zeta$ we may substitute in the
definition $\Phi_{\lambda}\left(  \Omega\right)  $ from Eq. \ref{phi},
multiplying by $R^{3N-1}G\left(  R\right)  ^{2}$ on both sides, integrating
over $R$ and using Eq. \ref{factoredint} to replace $V_{21_{R}}\left(
R\right)  V_{21_{\Omega}}\left(  \Omega\right)  $ gives a simple equation that
may be solve for $\zeta.$%
\begin{subequations}
\begin{align}
\zeta\alpha= &  \beta\tag{C3(a)}\label{constint2}\\
\alpha= &  \int R^{3N-1}G\left(  R\right)  ^{2}V_{21_{R}}\left(  R\right)
dR\tag{C3(b)}\label{alpha}\\
\beta= &  \dfrac{N\left(  \ N-1\right)  }{2}\int\prod\limits_{j=1}^{N}%
d^{3}rD^{\ast}\left(  \left\{  \vec{r}_{i}\right\}  _{i=1}^{N}\right)
\tag{C3(c)}\label{beta}\\
&  \times U_{int}\left(  \vec{r}_{21}\right)  D\left(  \left\{  \vec{r}%
_{i}\right\}  _{i=1}^{N}\right)  \nonumber
\end{align}
where $D\left(  \left\{  \vec{r}_{i}\right\}  _{i=1}^{N}\right)  $ is the
Slater determinant defined in Eq. \ref{Slater1} and $G\left(  R\right)  $ is
defined as in Eq. \ref{NIhyperrad}. We have also used that fact that
$R^{3N-1}dRd\Omega=\prod\limits_{j=1}^{N}d^{3}r_{j}$ \cite{Avery}. $\zeta$ can
now be found as a ratio of two integrals.

The integral, $\alpha$, on the LHS of Eq. \ref{constint2} can be calculated
directly. The integral, $\beta$, on the RHS of Eq. \ref{constint2} is now a
diagonal, determinantal matrix element. $\beta$ may be drastically simplified
by using the orthogonality of the single particle basis functions (for details
see \cite{cowan1981tas}, \S 6-1).%
\end{subequations}
\begin{align}
\beta=  &  \dfrac{1}{2}\sum_{i,j=1}^{N}\int d^{3}r_{1}d^{3}r_{2}%
\tag{C4}\label{DetermME}\\
&  \times\left[  \psi_{i}^{\ast}\left(  \vec{r}_{1}\right)  \psi_{j}^{\ast
}\left(  \vec{r}_{2}\right)  U_{int}\left(  \vec{r}_{21}\right)  \psi
_{i}\left(  \vec{r}_{1}\right)  \psi_{j}\left(  \vec{r}_{2}\right)  \right.
\nonumber\\
&  -\left.  \delta_{m_{s_{i}}m_{s_{j}}}\psi_{i}^{\ast}\left(  \vec{r}%
_{2}\right)  \psi_{j}^{\ast}\left(  \vec{r}_{1}\right)  U_{int}\left(  \vec
{r}_{21}\right)  \psi_{i}\left(  \vec{r}_{1}\right)  \psi_{j}\left(  \vec
{r}_{2}\right)  \right] \nonumber
\end{align}
Where $\psi_{i}\left(  \vec{r}\right)  $ is the $i$th single particle spatial
wave function that appears in the product in Eq. \ref{Slater1}, i.e.
\begin{equation}
r\psi_{i}\left(  \vec{r}\right)  =N_{n_{i}\ell_{i}}\exp\left(  -r^{2}%
/2l^{2}\right)  \left(  r/l\right)  ^{\ell_{i}+1}L_{n_{i}}^{\ell_{i}%
+1/2}\left[  \left(  r/l\right)  ^{2}\right]  y_{\ell_{i}m_{i}}\left(
\omega\right)  . \tag{C5}\label{spatial}%
\end{equation}

With this result we can now specify to the s-wave interaction. Following Eq.
\ref{factoredint} we may identify $U_{int}\left(  \vec{r}_{21}\right)
\longleftrightarrow g\delta^{3}\left(  \vec{r}_{21}\right)  $ and $V_{21_{R}%
}\left(  R\right)  \longleftrightarrow1/R^{3}$. With this and Eq.
\ref{NIhyperrad}, the integral in \ref{alpha} is found to be%
\begin{equation}
\alpha=\dfrac{\Gamma\left(  \lambda+\dfrac{3\left(  N-1\right)  }{2}\right)
}{\mathcal{L}^{3}\Gamma\left(  \lambda+\dfrac{3N}{2}\right)  }.\tag{C6}%
\label{alphaswave}%
\end{equation}
Evaluating Eq. \ref{DetermME} begins by integrating the $\delta$-function over
$\vec{r}_{2}$. This is simple and we can clearly see that the two terms in the
integral in Eq. \ref{DetermME} have a common factor of $\left\vert \psi
_{i}\left(  \vec{r}_{1}\right)  \right\vert ^{2}\left\vert \psi_{j}\left(
\vec{r}_{1}\right)  \right\vert ^{2}$. Factoring this out gives%
\begin{equation}
\beta=\dfrac{g}{2}\sum_{i,j=1}^{N}\left(  1-\delta_{m_{s_{i}}m_{s_{j}}%
}\right)  \int d^{3}r_{1}\left\vert \psi_{i}\left(  \vec{r}_{1}\right)
\right\vert ^{2}\left\vert \psi_{j}\left(  \vec{r}_{1}\right)  \right\vert
^{2}.\tag{C7}\label{betasimp}%
\end{equation}
This sum runs over all spatial and spin quantum numbers in a filled energy
shell, we may break this up into to factors, a spin sum and a space sum%
\begin{align}
\beta &  =\dfrac{g}{2}\sum_{m_{s_{i}},m_{s_{j}}=-1/2}^{1/2}\left(
1-\delta_{m_{s_{i}}m_{s_{j}}}\right)  \nonumber\\
&  \times\int d^{3}r_{1}\left(  \sum_{\nu}\left\vert \psi_{\nu}\left(  \vec
{r}_{1}\right)  \right\vert ^{2}\right)  \left(  \sum_{\mu}\left\vert
\psi_{\mu}\left(  \vec{r}_{1}\right)  \right\vert ^{2}\right)  \nonumber\\
&  =g\int d^{3}r_{1}\left(  \sum_{\nu}\left\vert \psi_{\nu}\left(  \vec{r}%
_{1}\right)  \right\vert ^{2}\right)  \left(  \sum_{\mu}\left\vert \psi_{\mu
}\left(  \vec{r}_{1}\right)  \right\vert ^{2}\right)  .\tag{C8}\label{betafin}%
\end{align}
where the Greek letters $\nu$ and $\mu$ stand for the set of spatial quantum
numbers $\left\{  n,\ell,m\right\}  $. The spin sum is trivial and is given by
$\sum_{m_{s_{i}},m_{s_{j}}=-1/2}^{1/2}\left(  1-\delta_{m_{s_{i}}m_{s_{j}}%
}\right)  =2$. We may now calculate $C_{N}$ for any given filled energy shell
using Eq. \ref{betafin} and plugging into the relationship
\begin{equation}
C_{N}=\dfrac{N^{3/2}}{g}\dfrac{\beta}{\alpha}.\tag{C9}\label{CNvsbeta}%
\end{equation}
This has been done for the first 100 filled shells the results of which are
summarized by Fig. \ref{CNfig}. $C_{N}$ for a selection of shells is given in
table \ref{Table1}.

To extract $C_{N}$ in the limit as $N\rightarrow\infty$ we may examine the
expression for $\beta$ in Eq. \ref{betafin}. We may note that the sum
$\rho\left(  \vec{r}\right)  =\sum_{\nu}\left\vert \psi_{\nu}\left(  \vec
{r}\right)  \right\vert ^{2}$ is the definition of the density of a spin
polarized degenerate fermi gas of non-interacting particles in an isotropic
oscillator. In the limit where $N\rightarrow\infty$ the trap energy of the
non-interacting gas dominates the total energy. This means that the
Thomas-Fermi approximation becomes exact in this limit. Thus we may write that%
\begin{align}
\rho\left(  \vec{r}\right)   &  =\dfrac{1}{6\pi^{2}}\left(  \dfrac{2m\mu
}{\hbar^{2}}\right)  ^{3/2}\left(  1-\dfrac{m\omega^{2}r^{2}}{2\mu}\right)
^{3/2}\tag{C10(a)}\label{dense}\\
N_{m_{s}}  &  =\int d^{3}r\rho\left(  \vec{r}\right)  \tag{C10(b)}%
\label{densecondition}%
\end{align}
The chemical potential $\mu$ may be found by the condition given in Eq.
\ref{densecondition} where $N_{m_{s}}$ is the number of particles with the
same spin projection $m_{s}$. The system we are considering is an equal spin
mixture so that $N_{\uparrow}=N_{\downarrow}=N/2$. Thus we find that
$\mu=\hbar\omega\left(  3N\right)  ^{1/3}$. Plugging this into Eq.
\ref{betafin} gives%
\begin{align}
\beta &  =g\int\left[  \rho\left(  r\right)  \right]  ^{2}d^{3}r\nonumber\\
&  =g\sqrt{\dfrac{2}{3}}\dfrac{256}{315}\dfrac{N^{3/2}}{\pi^{3}l^{3}}.
\tag{C11}\label{betabigN}%
\end{align}
We insert this into Eq. \ref{CNvsbeta} with Eq. \ref{alphaswave} to find that%
\begin{equation}
C_{N}=\sqrt{\dfrac{2}{3}}\dfrac{256}{315\pi^{3}}N^{3/2}\dfrac{\Gamma\left(
\lambda+\dfrac{3N}{2}\right)  }{\Gamma\left(  \lambda+\dfrac{3\left(
N-1\right)  }{2}\right)  } \tag{C12}\label{CNbig}%
\end{equation}
Using Eq. \ref{lambdavsN} for the large $N$ behavior of $\lambda$, in the
limit $N\rightarrow\infty$
\begin{equation}
C_{N}\longrightarrow\sqrt{\dfrac{2}{3}}\dfrac{32}{35\pi^{3}}N^{7/2}
\tag{C13}\label{CNfinal}%
\end{equation}
which is the quoted large $N$ behavior in Eq. \ref{CNlargeN}.

It should be said that the same formalism that is presented in this paper can
be applied to this system in center of mass coordinates. This is done by first
choosing an appropriate set of Jacobi coordinates. $\vec{x}_{i}=\sqrt
{i/\left(  i+1\right)  }\left(  \sum_{j=1}^{i}\vec{r}_{j}/i-r_{i+1}\right)  $
for $i=1,...N-1$ with the center of mass vector defined as $\vec{x}_{cm}%
=\sum_{j=1}^{N}\vec{r}_{j}/\sqrt{N}$. Hyperangular coordinates are now used to
describe the $3N-3$ degrees of freedom in the Jacobi coordinates where
$R^{2}=\sum_{j=1}^{N-1}x_{j}^{2}/N=\left(  \sum_{j=1}^{N}r_{j}^{2}-x_{cm}%
^{2}\right)  /N$. The $3N-4$ hyperangles needed are now defined with respect
to the lengths of the Jacobi vectors in the same way as in Eqs. \ref{hypera}
and \ref{hypera2}. Under this coordinate transformation we find the
Hamiltonian to be given by%
\[
H=H_{CM}+H_{R,\Omega}%
\]
Where $H_{CM}$ is the hamiltonian for the center of mass coordinate $\vec
{x}_{cm}$ and $H_{R,\Omega}$ is an operator entirely defined by hyperspherical
coordinates, i.e.%
\begin{align*}
&  H_{CM}=\dfrac{-\hbar^{2}}{2m}\nabla_{cm}^{2}+\dfrac{1}{2}m\omega^{2}%
x_{cm}^{2}\\
&  H_{R,\Omega}=\dfrac{-\hbar^{2}}{2M}\left(  \dfrac{1}{R^{3N-4}}%
\dfrac{\partial}{\partial R}R^{3N-4}\dfrac{\partial}{\partial R}%
-\dfrac{\mathbf{\Lambda}^{2}}{R^{2}}\right)  +\dfrac{1}{2}M\omega^{2}%
R^{2}+\sum_{i>j}U_{int}\left(  \vec{r}_{ij}\right)
\end{align*}
With this Hamiltonian we make the ansatz that for $H\Psi=E\Psi$, $\Psi
=\chi\left(  \vec{x}_{cm}\right)  G\left(  R\right)  \Phi_{\lambda}\left(
\Omega\right)  $ where $\chi$ is a wave function describing the center of mass
motion, $G\left(  R\right)  $ is a nodeless hyperradial function and
$\Phi_{\lambda}\left(  \Omega\right)  $ is the lowest hyperspherical harmonic
in the $3N-4$ angular coordinates. We are then looking for the matrix element
$\left\langle \Phi_{\lambda}\left\vert H_{R,\Omega}\right\vert \Phi_{\lambda
}\right\rangle $ where the integral is taken over all hyperangles at fixed
$R$. Theorem 1 still applies with the added idea that $\chi$ is given by the
lowest s-wave state of the center of mass in an oscillator. From here the
analysis presented in this section for trap centered coordinates still holds
with the added change that the dimension of the hyperradial integral in Eq.
\ref{alpha} is three dimensions smaller. This leads to a factor in the
interaction matrix element given by%
\[
C_{N}\rightarrow\dfrac{\left[  \Gamma\left(  \lambda+\dfrac{3\left(
N-1\right)  }{2}\right)  \right]  ^{2}}{\Gamma\left(  \lambda+\dfrac{3N}%
{2}\right)  \Gamma\left(  \lambda+\dfrac{3\left(  N-2\right)  }{2}\right)
}C_{N}.
\]
For smaller $N$ this factor changes the interaction considerably, but for
larger $N$ it quickly goes to $1$. Thus, in the large $N$ from limit, besides
extracting the center of mass \textquotedblleft sloshing" modes, there is very
little difference from the trap center coordinate systems in this Jacobi
coordinate formalism.

\section{APPENDIX D: CALCULATING $\rho\left(  0\right)  $}

We start from Eq. \ref{dense1}
\[
\rho\left(  0\right)  =N\int dRR^{3N-1}\left\vert F\left(  R\right)
\right\vert ^{2}\int d\Omega\delta^{3}\left(  \vec{r}_{N}\right)  \left\vert
\Phi_{\lambda}\left(  \Omega\right)  \right\vert ^{2}.
\]
If we can find $\int d\Omega\delta^{3}\left(  \vec{r}_{N}\right)  \left\vert
\Phi_{\lambda}\left(  \Omega\right)  \right\vert ^{2}$ then we have the
solution. From Eq. \ref{hypera2} and properties of the $\delta$-function we
may say that
\[
N\int d\Omega\delta^{3}\left(  \vec{r}_{N}\right)  \left\vert \Phi_{\lambda
}\left(  \Omega\right)  \right\vert ^{2}=\dfrac{\xi}{R^{3}}.
\]
We multiply this on both sides by the non-interacting hyperradial function
$R^{3N-1}\left(  G\left(  R\right)  \right)  ^{2}$ and integrate over $R$.%
\begin{equation}
N\int\int\delta^{3}\left(  \vec{r}_{N}\right)  \left\vert \Phi_{\lambda
}\left(  \Omega\right)  \right\vert ^{2}\left(  G\left(  R\right)  \right)
^{2}R^{3N-1}dRd\Omega=\xi\int\dfrac{\left(  G\left(  R\right)  \right)  ^{2}%
}{R^{3}}R^{3N-1}dR\tag{D1}\label{rhoint}%
\end{equation}
Inserting $G\left(  R\right)  $ from Eq. \ref{NIhyperrad} to the right hand
side is gives%
\begin{equation}
\xi\int\dfrac{\left(  G\left(  R\right)  \right)  ^{2}}{N^{3/2}R^{3}}%
R^{3N-1}dR=\xi\dfrac{\Gamma\left[  \lambda-3\left(  N-1\right)  /2\right]
}{\mathcal{L}^{3}\Gamma\left[  \lambda-3N/2\right]  }.\tag{D2(a)}\label{F1RHS}%
\end{equation}
From the definition of $\Phi_{\lambda}$ in Eq. \ref{phi} we see that the left
hand side of \ref{rhoint} is a determinantal matrix element of a single
particle operator integrated over all independent particle coordinates.
\begin{equation}
N\int\int\delta^{3}\left(  \vec{r}_{N}\right)  \left\vert \Phi_{\lambda
}\left(  \Omega\right)  \right\vert ^{2}\left(  G\left(  R\right)  \right)
^{2}R^{3N-1}dRd\Omega=N\int\prod\limits_{j=1}^{N}d^{3}r\left\vert D\left(
\left\{  \vec{r}_{i}\right\}  _{i=1}^{N}\right)  \right\vert ^{2}\delta
^{3}\left(  \vec{r}_{N}\right)  \tag{D2(b)}\label{F1LHS}%
\end{equation}
Where $D\left(  \left\{  \vec{r}_{i}\right\}  _{i=1}^{N}\right)  $ is the
Slater determinant wave function defined in Eq. \ref{Slater1}. Referring to
the definition of the density $\rho$ in Eq. \ref{densedef} we see that this is
merely the peak density of the non-interacting system $\rho_{NI}\left(
0\right)  $. Thus%
\[
\xi=\rho_{NI}\left(  0\right)  \dfrac{l^{3}\Gamma\left[  \lambda-3N/2\right]
}{N^{3/2}\Gamma\left[  \lambda-3\left(  N-1\right)  /2\right]  }.
\]
The peak density is then given by%
\[
\rho\left(  0\right)  =\xi\int dR\dfrac{R^{3N-1}\left\vert F\left(  R\right)
\right\vert ^{2}}{R^{3}}%
\]
which is what we were seeking to show.


\begin{thebibliography}{25}
\expandafter\ifx\csname natexlab\endcsname\relax\def\natexlab#1{#1}\fi
\expandafter\ifx\csname bibnamefont\endcsname\relax
  \def\bibnamefont#1{#1}\fi
\expandafter\ifx\csname bibfnamefont\endcsname\relax
  \def\bibfnamefont#1{#1}\fi
\expandafter\ifx\csname citenamefont\endcsname\relax
  \def\citenamefont#1{#1}\fi
\expandafter\ifx\csname url\endcsname\relax
  \def\url#1{\texttt{#1}}\fi
\expandafter\ifx\csname urlprefix\endcsname\relax\def\urlprefix{URL }\fi
\providecommand{\bibinfo}[2]{#2}
\providecommand{\eprint}[2][]{\url{#2}}

\bibitem[{\citenamefont{Bartenstein et~al.}(2004)\citenamefont{Bartenstein,
  Altmeyer, Riedl, Jochim, Chin, Denschlag, and Grimm}}]{bartenstein2004cmb}
\bibinfo{author}{\bibfnamefont{M.}~\bibnamefont{Bartenstein}},
  \bibinfo{author}{\bibfnamefont{A.}~\bibnamefont{Altmeyer}},
  \bibinfo{author}{\bibfnamefont{S.}~\bibnamefont{Riedl}},
  \bibinfo{author}{\bibfnamefont{S.}~\bibnamefont{Jochim}},
  \bibinfo{author}{\bibfnamefont{C.}~\bibnamefont{Chin}},
  \bibinfo{author}{\bibfnamefont{J.~H.}~\bibnamefont{Denschlag}},
  \bibnamefont{and} \bibinfo{author}{\bibfnamefont{R.}~\bibnamefont{Grimm}},
  \bibinfo{journal}{Phys. Rev. Lett.} \textbf{\bibinfo{volume}{92}},
  \bibinfo{pages}{120401} (\bibinfo{year}{2004}).

\bibitem[{\citenamefont{Bourdel et~al.}(2004)\citenamefont{Bourdel, Khaykovich,
  Cubizolles, Zhang, Chevy, Teichmann, Tarruell, Kokkelmans, and
  Salomon}}]{bourdel2004esb}
\bibinfo{author}{\bibfnamefont{T.}~\bibnamefont{Bourdel}},
  \bibinfo{author}{\bibfnamefont{L.}~\bibnamefont{Khaykovich}},
  \bibinfo{author}{\bibfnamefont{J.}~\bibnamefont{Cubizolles}},
  \bibinfo{author}{\bibfnamefont{J.}~\bibnamefont{Zhang}},
  \bibinfo{author}{\bibfnamefont{F.}~\bibnamefont{Chevy}},
  \bibinfo{author}{\bibfnamefont{M.}~\bibnamefont{Teichmann}},
  \bibinfo{author}{\bibfnamefont{L.}~\bibnamefont{Tarruell}},
  \bibinfo{author}{\bibfnamefont{S.~J.~J.~M.~F.}~\bibnamefont{Kokkelmans}},
  \bibnamefont{and} \bibinfo{author}{\bibfnamefont{C.}~\bibnamefont{Salomon}},
  \bibinfo{journal}{Phys. Rev. Lett.} \textbf{\bibinfo{volume}{93}},
  \bibinfo{pages}{50401} (\bibinfo{year}{2004}).

\bibitem[{\citenamefont{Kinast et~al.}(2004)\citenamefont{Kinast, Hemmer, Gehm,
  Turlapov, and Thomas}}]{kinast2004esr}
\bibinfo{author}{\bibfnamefont{J.}~\bibnamefont{Kinast}},
  \bibinfo{author}{\bibfnamefont{S.~L.}~\bibnamefont{Hemmer}},
  \bibinfo{author}{\bibfnamefont{M.~E.}~\bibnamefont{Gehm}},
  \bibinfo{author}{\bibfnamefont{A.}~\bibnamefont{Turlapov}}, \bibnamefont{and}
  \bibinfo{author}{\bibfnamefont{J.~E.}~\bibnamefont{Thomas}},
  \bibinfo{journal}{Physical Review Letters} \textbf{\bibinfo{volume}{92}},
  \bibinfo{pages}{150402} (\bibinfo{year}{2004}).

\bibitem[{\citenamefont{Regal et~al.}(2004)\citenamefont{Regal, Greiner, and
  Jin}}]{regal2004orc}
\bibinfo{author}{\bibfnamefont{C.~A.}~\bibnamefont{Regal}},
  \bibinfo{author}{\bibfnamefont{M.}~\bibnamefont{Greiner}}, \bibnamefont{and}
  \bibinfo{author}{\bibfnamefont{D.~S.}~\bibnamefont{Jin}},
  \bibinfo{journal}{Phys. Rev. Lett.} \textbf{\bibinfo{volume}{92}},
  \bibinfo{pages}{40403} (\bibinfo{year}{2004}).

\bibitem[{\citenamefont{Zwierlein et~al.}(2004)\citenamefont{Zwierlein, Stan,
  Schunck, Raupach, Kerman, and Ketterle}}]{zwierlein2004cpf}
\bibinfo{author}{\bibfnamefont{M.~W.}~\bibnamefont{Zwierlein}},
  \bibinfo{author}{\bibfnamefont{C.~A.}~\bibnamefont{Stan}},
  \bibinfo{author}{\bibfnamefont{C.~H.}~\bibnamefont{Schunck}},
  \bibinfo{author}{\bibfnamefont{S.~M.~F.}~\bibnamefont{Raupach}},
  \bibinfo{author}{\bibfnamefont{A.~J.}~\bibnamefont{Kerman}}, \bibnamefont{and}
  \bibinfo{author}{\bibfnamefont{W.}~\bibnamefont{Ketterle}},
  \bibinfo{journal}{Phys. Rev. Lett.} \textbf{\bibinfo{volume}{92}},
  \bibinfo{pages}{120403} (\bibinfo{year}{2004}).

\bibitem[{\citenamefont{Rittenhouse et~al.}(2005)\citenamefont{Rittenhouse,
  Cavagnero, von Stecher, and Greene}}]{condmat}
\bibinfo{author}{\bibfnamefont{S.~T.} \bibnamefont{Rittenhouse}},
  \bibinfo{author}{\bibfnamefont{M.~J.} \bibnamefont{Cavagnero}},
  \bibinfo{author}{\bibfnamefont{J.}~\bibnamefont{von Stecher}},
  \bibnamefont{and} \bibinfo{author}{\bibfnamefont{C.~H.}
  \bibnamefont{Greene}}, \bibinfo{journal}{arXiv:cond-mat/0510454}
  (\bibinfo{year}{2005}).

\bibitem[{\citenamefont{Bohn et~al.}(1998)\citenamefont{Bohn, Esry, and
  Greene}}]{bohn_esry_greene_hsbec}
\bibinfo{author}{\bibfnamefont{J.~L.} \bibnamefont{Bohn}},
  \bibinfo{author}{\bibfnamefont{B.~D.} \bibnamefont{Esry}}, \bibnamefont{and}
  \bibinfo{author}{\bibfnamefont{C.~H.} \bibnamefont{Greene}},
  \bibinfo{journal}{Phys. Rev. A} \textbf{\bibinfo{volume}{58}},
  \bibinfo{pages}{584} (\bibinfo{year}{1998}).

\bibitem[{\citenamefont{Kim and Zubarev}(2000)}]{kim2000elt}
\bibinfo{author}{\bibfnamefont{Y.~E.} \bibnamefont{Kim}} \bibnamefont{and}
  \bibinfo{author}{\bibfnamefont{A.}~\bibnamefont{Zubarev}},
  \bibinfo{journal}{J. Phys. B} \textbf{\bibinfo{volume}{33}},
  \bibinfo{pages}{55} (\bibinfo{year}{2000}).

\bibitem[{\citenamefont{Kushibe et~al.}(2004)\citenamefont{Kushibe, Mutou,
  Morishita, Watanabe, and Matsuzawa}}]{kushibe2004aha}
\bibinfo{author}{\bibfnamefont{D.}~\bibnamefont{Kushibe}},
  \bibinfo{author}{\bibfnamefont{M.}~\bibnamefont{Mutou}},
  \bibinfo{author}{\bibfnamefont{T.}~\bibnamefont{Morishita}},
  \bibinfo{author}{\bibfnamefont{S.}~\bibnamefont{Watanabe}}, \bibnamefont{and}
  \bibinfo{author}{\bibfnamefont{M.}~\bibnamefont{Matsuzawa}},
  \bibinfo{journal}{Phys. Rev. A} \textbf{\bibinfo{volume}{70}},
  \bibinfo{pages}{63617} (\bibinfo{year}{2004}).

\bibitem[{\citenamefont{Smirnov and Shitikova}(1977)}]{SmirnovShitikova}
\bibinfo{author}{\bibfnamefont{Y.~F.} \bibnamefont{Smirnov}} \bibnamefont{and}
  \bibinfo{author}{\bibfnamefont{K.~V.} \bibnamefont{Shitikova}},
  \bibinfo{journal}{Sov. J. Part. Nucl.} \textbf{\bibinfo{volume}{8}},
  \bibinfo{pages}{44} (\bibinfo{year}{1977}).

\bibitem[{\citenamefont{Blume and Greene}(2000)}]{blume00}
\bibinfo{author}{\bibfnamefont{D.}~\bibnamefont{Blume}} \bibnamefont{and}
  \bibinfo{author}{\bibfnamefont{C.~H.} \bibnamefont{Greene}},
  \bibinfo{journal}{J. Chem. Phys.} \textbf{\bibinfo{volume}{112}},
  \bibinfo{pages}{8053} (\bibinfo{year}{2000}).

\bibitem[{\citenamefont{Esry et~al.}(1999)\citenamefont{Esry, Greene, and
  Jr}}]{esry1999rta}
\bibinfo{author}{\bibfnamefont{B.~D.} \bibnamefont{Esry}},
  \bibinfo{author}{\bibfnamefont{C.~H.} \bibnamefont{Greene}},
  \bibnamefont{and} \bibinfo{author}{\bibfnamefont{J.~P.} \bibnamefont{Burke}},
  \bibinfo{journal}{Phys. Rev. Lett.} \textbf{\bibinfo{volume}{83}},
  \bibinfo{pages}{1751} (\bibinfo{year}{1999}).

\bibitem[{\citenamefont{Avery}(1989)}]{Avery}
\bibinfo{author}{\bibfnamefont{J.}~\bibnamefont{Avery}},
  \emph{\bibinfo{title}{{Hyperspherical Harmonics: Applications in Quantum
  Theory}}} (\bibinfo{publisher}{Kluwer Academic Publishers},
  \bibinfo{year}{1989}).

\bibitem[{\citenamefont{Barnea}(1999)}]{barnea1999rmc}
\bibinfo{author}{\bibfnamefont{N.}~\bibnamefont{Barnea}}, \bibinfo{journal}{J.
  Math. Phys.} \textbf{\bibinfo{volume}{40}}, \bibinfo{pages}{1011}
  (\bibinfo{year}{1999}).

\bibitem[{\citenamefont{Cavagnero}(1986)}]{Cavagnero86}
\bibinfo{author}{\bibfnamefont{M.~J.} \bibnamefont{Cavagnero}},
  \bibinfo{journal}{Phys. Rev. A} \textbf{\bibinfo{volume}{33}},
  \bibinfo{pages}{2877} (\bibinfo{year}{1986}).

\bibitem[{\citenamefont{Fabre de~la Ripelle and Navarro}(1978)}]{Fabre78}
\bibinfo{author}{\bibfnamefont{M.}~\bibnamefont{Fabre de~la Ripelle}}
  \bibnamefont{and} \bibinfo{author}{\bibfnamefont{J.}~\bibnamefont{Navarro}},
  \bibinfo{journal}{Ann. Phys.} \textbf{\bibinfo{volume}{123}},
  \bibinfo{pages}{185} (\bibinfo{year}{1978}).

\bibitem[{\citenamefont{Fabre de~la Ripelle et~al.}(2005)\citenamefont{Fabre
  de~la Ripelle, Sofianos, and Adam}}]{Fabre05}
\bibinfo{author}{\bibfnamefont{M.}~\bibnamefont{Fabre de~la Ripelle}},
  \bibinfo{author}{\bibfnamefont{S.~A.} \bibnamefont{Sofianos}},
  \bibnamefont{and} \bibinfo{author}{\bibfnamefont{R.~M.} \bibnamefont{Adam}},
  \bibinfo{journal}{Ann. Phys.} \textbf{\bibinfo{volume}{316}},
  \bibinfo{pages}{107} (\bibinfo{year}{2005}).

\bibitem[{\citenamefont{Timofeyuk}(2002)}]{Timofeyuk02}
\bibinfo{author}{\bibfnamefont{N.~K.} \bibnamefont{Timofeyuk}},
  \bibinfo{journal}{Phys. Rev. C} \textbf{\bibinfo{volume}{65}},
  \bibinfo{pages}{064306} (\bibinfo{year}{2002}).

\bibitem[{\citenamefont{Timofeyuk}(2004)}]{Timofeyuk04}
\bibinfo{author}{\bibfnamefont{N.~K.} \bibnamefont{Timofeyuk}},
  \bibinfo{journal}{Phys. Rev. C} \textbf{\bibinfo{volume}{69}},
  \bibinfo{pages}{034336} (\bibinfo{year}{2004}).

\bibitem[{\citenamefont{Fermi}(1934)}]{fermi1934}
\bibinfo{author}{\bibfnamefont{E.}~\bibnamefont{Fermi}},
  \bibinfo{journal}{Nuovo Cimento} \textbf{\bibinfo{volume}{11}},
  \bibinfo{pages}{157} (\bibinfo{year}{1934}).

\bibitem[{\citenamefont{Esry and Greene}(1999)}]{esry1999vsi}
\bibinfo{author}{\bibfnamefont{B.~D.} \bibnamefont{Esry}} \bibnamefont{and}
  \bibinfo{author}{\bibfnamefont{C.~H.} \bibnamefont{Greene}},
  \bibinfo{journal}{Phys. Rev. A} \textbf{\bibinfo{volume}{60}},
  \bibinfo{pages}{1451} (\bibinfo{year}{1999}).

\bibitem[{\citenamefont{Vichi and Stringari}(1999)}]{vichi1999coi}
\bibinfo{author}{\bibfnamefont{L.}~\bibnamefont{Vichi}} \bibnamefont{and}
  \bibinfo{author}{\bibfnamefont{S.}~\bibnamefont{Stringari}},
  \bibinfo{journal}{Phys. Rev. A} \textbf{\bibinfo{volume}{60}},
  \bibinfo{pages}{4734} (\bibinfo{year}{1999}).

\bibitem[{\citenamefont{Germann et~al.}(1995)\citenamefont{Germann, Herschbach,
  Dunn, and Watson}}]{Germann95}
\bibinfo{author}{\bibfnamefont{T.~C.} \bibnamefont{Germann}},
  \bibinfo{author}{\bibfnamefont{D.~R.} \bibnamefont{Herschbach}},
  \bibinfo{author}{\bibfnamefont{M.}~\bibnamefont{Dunn}}, \bibnamefont{and}
  \bibinfo{author}{\bibfnamefont{D.~K.} \bibnamefont{Watson}},
  \bibinfo{journal}{Phys. Rev. Lett.} \textbf{\bibinfo{volume}{74}},
  \bibinfo{pages}{658} (\bibinfo{year}{1995}).

\bibitem[{\citenamefont{Fregoso and Baym}(2006)}]{Fregoso06}
\bibinfo{author}{\bibfnamefont{B.~M.} \bibnamefont{Fregoso}} \bibnamefont{and}
  \bibinfo{author}{\bibfnamefont{G.}~\bibnamefont{Baym}},
  \bibinfo{journal}{arXiv:cond-mat/0602191}  (\bibinfo{year}{2006}).

\bibitem[{\citenamefont{Cowan}(1981)}]{cowan1981tas}
\bibinfo{author}{\bibfnamefont{R.~D.} \bibnamefont{Cowan}},
  \emph{\bibinfo{title}{The Theory of Atomic Structure and Spectra}}
  (\bibinfo{publisher}{University of California Press}, \bibinfo{year}{1981}).

\end{thebibliography}
\end{document}